%% file: main.tex
\documentclass[a4paper,11pt,elsarticle]{article}
\pdfoutput=1
\usepackage{jinstpub}
\usepackage[normalem]{ulem}
\usepackage[symbol]{footmisc}
\usepackage{xcolor}
\usepackage{graphicx}
\usepackage{mathtools}
\usepackage{comment}
\usepackage{tabularx}
\usepackage[figuresright]{rotating}
\usepackage{upgreek}
\usepackage{amsmath,graphicx} 
\usepackage{multirow}
\usepackage{color}
\usepackage{multirow}
\usepackage{scrextend}
\usepackage{tabu}
\usepackage{color, colortbl}
\usepackage{soul}
\usepackage{babel,blindtext}
\usepackage{caption}
\usepackage{subcaption}
\newcolumntype{P}[1]{>{\centering\arraybackslash}p{#1}}
\definecolor{LightCyan}{rgb}{0.88,1,1}
\usepackage{hyperref} 
 \hypersetup{ 
     colorlinks=true, 
     linkcolor=black, 
     filecolor=black, 
     citecolor = black,       
     urlcolor=black, 
     } 
     
\title{\boldmath The Design, Implementation, and Performance of the LZ Calibration Systems}

\input{LZ_Latest_Author_List-copy_to_Overleaf_Elsevier_April24_2024.tex}

\abstract{
LUX-ZEPLIN (LZ) is a tonne-scale experiment searching for direct dark matter interactions and other rare events. It is located at the Sanford Underground Research Facility (SURF) in Lead, South Dakota, USA. The core of the LZ detector is a dual-phase xenon time projection chamber (TPC), designed with the primary goal of detecting Weakly Interacting Massive Particles (WIMPs) via their induced low energy nuclear recoils. Surrounding the TPC, two veto detectors immersed in an ultra-pure water tank enable reducing background events to enhance the discovery potential. Intricate calibration systems are purposely designed to precisely understand the responses of these three detector volumes to various types of particle interactions and to demonstrate LZ's ability to discriminate between signals and backgrounds. In this paper, we present a comprehensive discussion of the key features, requirements, and performance of the LZ calibration systems, which play a crucial role in enabling LZ's WIMP-search and its broad science program. The thorough description of these calibration systems, with an emphasis on their novel aspects, is valuable for future calibration efforts in direct dark matter and other rare-event search experiments.
}

\collaboration{The LZ Collaboration}
\begin{document}\sloppy
\maketitle
\flushbottom
\section{Introduction}\label{intro}
The nature of dark matter remains one of the biggest mysteries of modern physics. Despite  multiple astronomical and cosmological observations indicating that dark matter constitutes ${\sim 85\%}$ of the total mass in the universe  \cite{doi:10.1146/annurev.astro.39.1.137,Planck:2018vyg,BOSS:2013rlg,Clowe:2006eq,Hu:2001bc}, it has not been directly detected so far. Weakly Interacting Massive Particles (WIMPs), arising from various theories beyond the Standard Model \cite{RevModPhys.90.045002,Billard:2021uyg,Akerib:2022ort}, remain one of the most promising dark matter particle candidates \cite{arcadi2018waning,PhysRevLett.39.165,bertone2005particle}. These WIMPs, along with other dark matter particle candidates~\cite{barnes2020simple,das2022mini}, are searched for with colliders, telescopes and underground-based experiments using different approaches and detection technologies~\cite{Akerib:2022ort,roszkowski2018wimp}. 

The LUX-ZEPLIN (LZ) experiment utilizes a dual-phase xenon time projection chamber (TPC) technology aided by two surrounding veto detectors to reject numerous backgrounds while primarily looking for low energy nuclear recoil signals from dark matter particles interacting inside the detector. 
Xenon recoils from particle interactions in the TPC yield two signatures: the emission of a prompt light signal followed by a delayed/secondary scintillation signal produced by charge extracted from the interaction site. In order to make an unambiguous WIMP dark matter detection or set stringent limits on their interactions in LZ, a rigorous calibration system is implemented to understand the entire range of predicted WIMP signatures (light and charge) and that of their backgrounds. The responses of the surrounding veto volumes to all particle interactions are also carefully studied to optimize the background veto efficiency of these detectors and to enhance the TPC discovery capability.

In addition to detecting WIMP dark matter through their nuclear recoils (NR) in xenon~\cite{LZ:2022ufs}, the LZ TPC detector can also be used for other rare-event searches~\cite{aalbers2024new, mount2017lux}. These include observation of $^{8}$B solar neutrinos through low energy coherent elastic neutrino-nucleus scattering (CE$\nu$NS) signals, measurements of electromagnetic properties of solar neutrinos, and detection of other dark matter candidates such as solar axions and dark photons via low energy electronic recoil (ER) signals~\cite{aalbers2023search}. The broad LZ science program also encompasses measuring effective field theory NR couplings for dark matter~\cite{aalbers2023first}, searching for neutrinoless double-beta decay ($0\nu\beta\beta$) of $^{136}$Xe~\cite{collaboration2020projected}, and looking for rare decays of other xenon isotopes. Therefore, the TPC response to nuclear recoils of energy ranging from $<$2~keV$_{nr}$\footnote{nuclear recoil equivalent energy}~\cite{ma2023search, aprile2021search} to $\sim$240~keV$_{nr}$~\cite{XENON:2017fdd,akerib2021constraints} and electronic recoils of energy beyond the $Q_{\beta \beta}$ of $^{136}$Xe = 2,458~keV$_{ee}$\footnote{electron equivalent recoil energy}~\cite{redshaw2007mass,collaboration2020projected} needs to be characterized using the LZ calibration systems.

This paper is a comprehensive documentation of the source calibration systems for LZ, including the science goals, design, implementation, and performance of each calibration system. In section~\ref{overview}, an overview of the LZ detectors and their calibration systems is presented.  This section also describes a list of calibration sources that will be used throughout the entirety of LZ operation. Sections \ref{disbursed} - \ref{ybe} provide detailed descriptions of the functionality and performance of the sources and the calibration systems: the dispersed source injection (SI) system, the external rod calibration source deployment (CSD) system, the deuterium-deuterium (DD) neutron generator system, and the photo-neutron (YBe) delivery system. Section~\ref{pmtcalib} describes the calibrations of LZ's photomultiplier tubes (PMTs) in the TPC and veto detectors using light-emitting diodes (LEDs) for monitoring their respective gains and the stability of their light response over time. 

\begin{figure}[ht]
  \centering
  \includegraphics[width=0.8\textwidth]{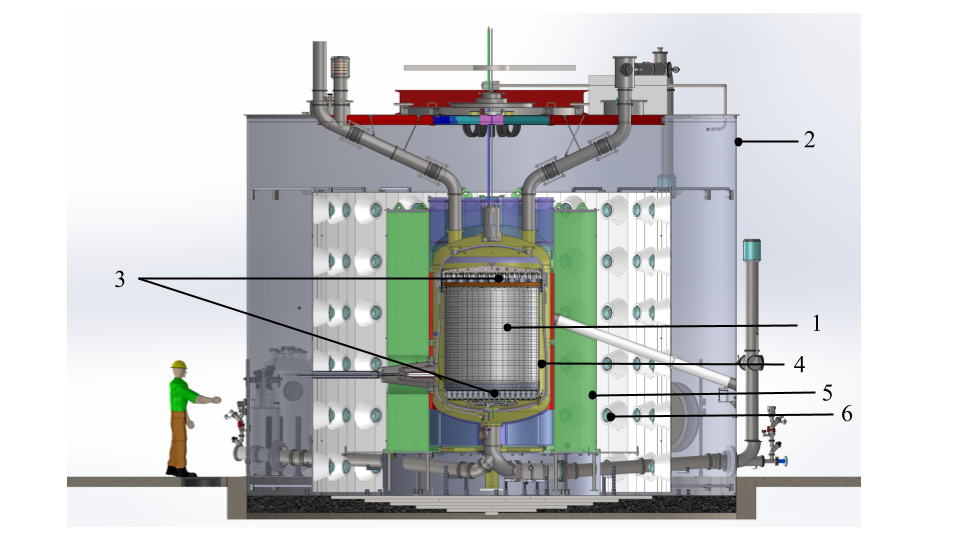}
  \caption{Overview of LZ detectors consisting of the central TPC detector (1) surrounded by the Skin detector (4) and the OD (5) volumes. The OD is filled with gadolinium-loaded liquid scintillator and viewed by OD PMTs (6). All three detectors are enclosed in a water tank (2) filled with high purity water. The PMT arrays (3) monitoring the TPC xenon volume are also shown.
  }
  \label{fig:overview}
\end{figure}

\section{Overview of the LZ Detectors and of the Calibration Systems} \label{overview}

LZ is a low-background experiment~\cite{LZ:2022ysc} located 4,850 feet underground at the Sanford Underground Research Facility (SURF)~\cite{heise2020sanford} to benefit from the rock overburden for cosmogenic background reduction. The experiment is optimized to look for WIMPs with masses above $\sim$10~GeV/c$^2$~\cite{akerib2015lux,mount2017lux,collaboration2020lux}, as demonstrated by its first results~\cite{LZ:2022ufs}. Its detectors are organized in a nested structure as shown in Figure~\ref{fig:overview}. The central TPC is a cylindrical barrel of approximately 1.5~m diameter and 1.5~m height, lined with reflective PTFE. The active volume, consisting of 7 tonnes of liquid xenon (LXe) contained between the gate and cathode electrodes and the PTFE wall, is viewed by 494 3-inch VUV sensitive PMTs installed in arrays at the top and bottom of the TPC~\cite{faham2015measurements}. The TPC is enclosed in a radio-pure double-walled titanium cryostat vessel~\cite{akerib2017identification,akerib2020lux} which is surrounded by the "Skin" and the outer detector (OD), both designed to provide veto signals for rejecting internal and external backgrounds. The Skin detector, containing $\sim$2 tonnes of LXe instrumented with 93 1-inch and 38 2-inch PMTs, is located between the outside of the TPC's PTFE walls and the inner cryostat vessel (ICV). The OD system is made of a set of acrylic tanks containing approximately 17 tonnes of gadolinium-loaded (0.1\% by mass) liquid scintillator (GdLS)~\cite{HASELSCHWARDT2019148}. These acrylic tanks were designed with custom cut-out holes to enable external calibration source deployment conduits to be as close to the TPC as possible. The entire detector assembly is located in a tank filled with 238 tonnes of ultra-pure water to mitigate against residual cosmogenic backgrounds and neutrons from the ambient environment. The water tank has 120 8-inch PMTs mounted on stainless steel frames to detect OD signals and Cherenkov light in the water. 

\begin{table}[ht]
\vspace{0.3cm}
\centering
\begin{tabular}
{|l|l|l|l|l|l|} 
\hline
 & Isotope & Particle species used & Energy (keV) & Half-life & Deployment\\
 & &   & &$t_{1/2}$ &\\ \hline
 \multirow{4}{*}& $^{83\mathrm{m}}$Kr ($^{83}$Rb) & IC, AE, x-ray, $\gamma$ & 32.1 and 9.4 &1.83~h& Internal \\ 
A& $^{131\mathrm{m}}$Xe ($^{131}$I) & $\gamma$, x-ray & {163.9}  &11.8~d&    \\
& $^{220}$Rn ($^{228}$Th)& $\alpha, \beta, \gamma$ & various~\cite{J_rg_2023} &55.6~s&\\ \hline
\multirow{3}{*}{B}& $^3$H & $\beta$ & {0 -- 18.6}  &12.3~y& Internal\\
& $^{14}$C & $\beta$ & {0 -- 156.4} &5730~y& \\
\hline
\multirow{9}{*}& $^{241}$AmLi & ($\alpha, n$) &  (5638, 0 $-$ 1500) &433~y& \\ 
& $^{241}$AmBe & ($\alpha, n$) & (5638, 0 $-$ $11\times10^3$) &433~y&  \\
& $^{57}$Co & $\gamma$ & {122} &272~d&External\\
C& $^{228}$Th & $\gamma$ & {2615} &1.91~y&\\
& $^{22}$Na & $\gamma$ & {511, 1275}  &2.60~y&\\
& $^{54}$Mn & $\gamma$ & {835}   &312~d&\\
\hline \multirow{4}{*}{} & $^{88}$YBe & ($\gamma, n$) & (1836, 152)  &107~d& External \\ 
D & $^{88}$YMg& $\gamma$& {1836}& 107~d& \\
\hline
 \multirow{3}{*}{E}& DD &  n & 2450 & $-$&  \\
& D-Reflector & n & 270 - 420  &$-$& {External}\\
& H-Reflector & n & 10 - 200 & $-$&   \\  \hline
\multirow{1}{*}{F}&$^{241}$Am &  $\alpha$ & {5638}  &$433$~y& External \\
\hline
\end{tabular}
\caption{A list of calibration sources used by LZ for the TPC, Skin, and OD detector calibrations. Category A isotopes are generator sources whose parent isotopes are shown in the brackets. The category C-E isotopes are used for calibrations of all three detectors while category A and B isotopes are only used for TPC calibrations. The category F isotope is used for calibrating the monitoring PMT located in the dark box of the OD Optical Calibration System. The abbreviations IC and AE refer to internal conversion electrons and Auger electrons produced by the $^{83\mathrm{m}}$Kr source. The energy (keV) refers to particle energies relevant for the calibration of LZ and is not a complete list of decay energies. The energies quoted in the parentheses correspond to those of the particle species from the previous column.}
\label{sourcesummary}
\end{table}

\begin{figure}[ht]
  \centering
  \includegraphics[width=0.9\textwidth]{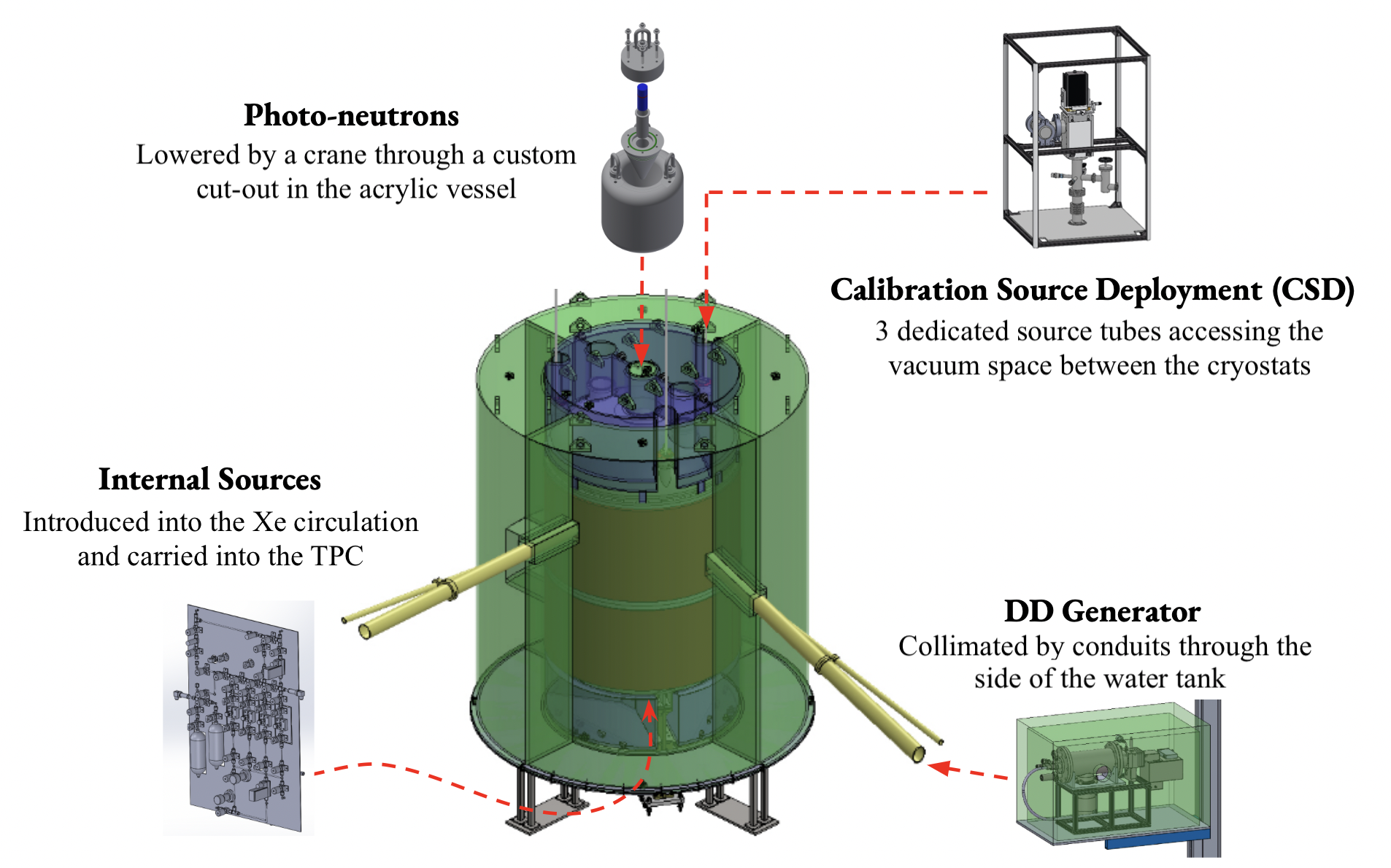}
  \caption{A schematic of the calibration deployment methods used for category A-E isotopes in Table~\ref{sourcesummary} (items are not shown to scale). Due to the variation in the physical size, form factors, and production mechanism of the calibration sources, different hardware and deployment methods are required.} \label{fig:dep-methods}
\end{figure}

Particles interacting in the active xenon TPC region can deposit a portion of their energy which is transferred into prompt scintillation light and ionization electrons. The prompt scintillation light can be detected within 100~ns and is referred to as S1. The ionization electrons drift under an applied uniform electric field to the liquid surface where they are extracted by a stronger electric field and produce a secondary scintillation in the xenon gas, called S2. Both the S1 and S2 light signals are detected by the PMTs in the top and bottom arrays. The ratio of S2 to S1 differentiates interactions with a xenon nucleus (producing a nuclear recoil) from interactions with the atomic electrons (producing an electronic recoil). 
A variety of calibration sources are used to understand the micro-physics of particle interactions inside the TPC, the Skin, and the OD, as well as the position and time dependence of these detector responses. Table~\ref{sourcesummary} summarizes these sources per category, their half-lives, their purposes and their deployment methods. The activities at procurement will be discussed for each source individually.
Figure~\ref{fig:dep-methods} shows the hardware and deployment methods for category A-E sources, designed according to the sources' physical size, form factors, and production mechanism. The category F consists of a sealed $^{241}$Am source inside a YAP:Ce crystal~\cite{YapCe}, which is used for calibrating the PMT in the dark box of the OD Optical Calibration System. Details about each calibration system will be discussed in the following sections.

\section{Dispersed Source Calibration }\label{disbursed}

\subsection{Xenon Circulation System}
 \label{circulation}
The LZ xenon circulation system is discussed in detail in reference~\cite{collaboration2020lux}. Only its relevant features to the dispersed/gaseous source calibration are highlighted here. The main feature is the continuous circulation of the xenon through a commercial hot zirconium getter (model PS5-MGT50-R from SAES) \cite{getter_model} to remove electronegative impurities. During Science Run 1 (SR1), with the LZ flow rate of $\sim$400~slpm (standard liters per minute), the entire xenon payload is transported through the getter every 3 days. The getter is also critical for the removal of methane-based radioactive gaseous sources after their deployment for electronic recoil calibrations, as will be discussed in section~\ref{injection}. As the purification takes place in the gas-phase, xenon has to be continuously evaporated and condensed. Condensation and evaporation happen in the liquid xenon tower, which features a heat exchanger for efficient liquid-gas interface handling. The circulation is driven by two all-metal diaphragm gas compressors (model A2-5/15 from Fluitron) \cite{compressor_model}. 
The purified and re-condensed xenon is delivered into the bottom of the ICV via two transfer lines. One feeds the TPC directly; the other allows flow into the Skin. Cryovalves allow the flow into these regions to be adjusted. At the top of the TPC, LXe spills over a set of six weirs. The spillover is collected in three drain lines which combine into one transfer line, returning the liquid in the circulation system for purification.

Gaseous sources for internal calibrations can be injected/pushed into the circulation flow path before or after the getter, as required by the source type. They then flow into the main circulation system and enter the TPC at controlled dose via seven LXe inlet ports beneath the TPC, as will be detailed below.

\subsection{Dispersed Sources}
\label{disbursed_sources}
LZ uses several gaseous calibration radioisotopes which are injected (pushed by flowing high pressure xenon gas) into the main xenon circulation system. The circulation flow then carries the injected radioactivity to the condensing stage, and finally with the LXe into the TPC volume itself. This deployment of calibration isotopes directly into the TPC volume is desired for several reasons, including 1) overcoming the self-shielding of the LXe target to external radiation; 2) providing a calibration with a spatial uniformity; 3) observing how injected activities mix within the TPC to improve understanding of background radioisotope flow and mixing.  The five dispersed sources used in LZ are listed in Table~\ref{sourcesummary} as categories A and B.  The $^3$H and $^{14}$C isotopes are stored in pressurized cylinders as either $^3$H- or $^{14}$C-labeled CH$_4$ mixed with Xe carrier gas. The $^{83\mathrm{m}}$Kr, $^{131\mathrm{m}}$Xe, and $^{220}$Rn isotopes are produced by progenitor nuclei, which are solid materials that decay into these species. These latter three are referred to as ``generator'' sources, while the former two are referred to as ``bottle'' sources.

The generator sources must completely retain the parent isotope (to prevent a long-lived isotope contaminating the plumbing of either the injection panel or the circulation system) while allowing efficient transport of the gaseous daughter calibration isotope.  In each of the generator sources, the daughter is a noble element, aiding its escape from the parent materials, and allowing the calibration gas to be injected into the circulation path upstream of the getter (which allows noble radioisotopes through).  

\begin{figure}[ht]
\centering
  \centering
  \includegraphics[width=\textwidth]{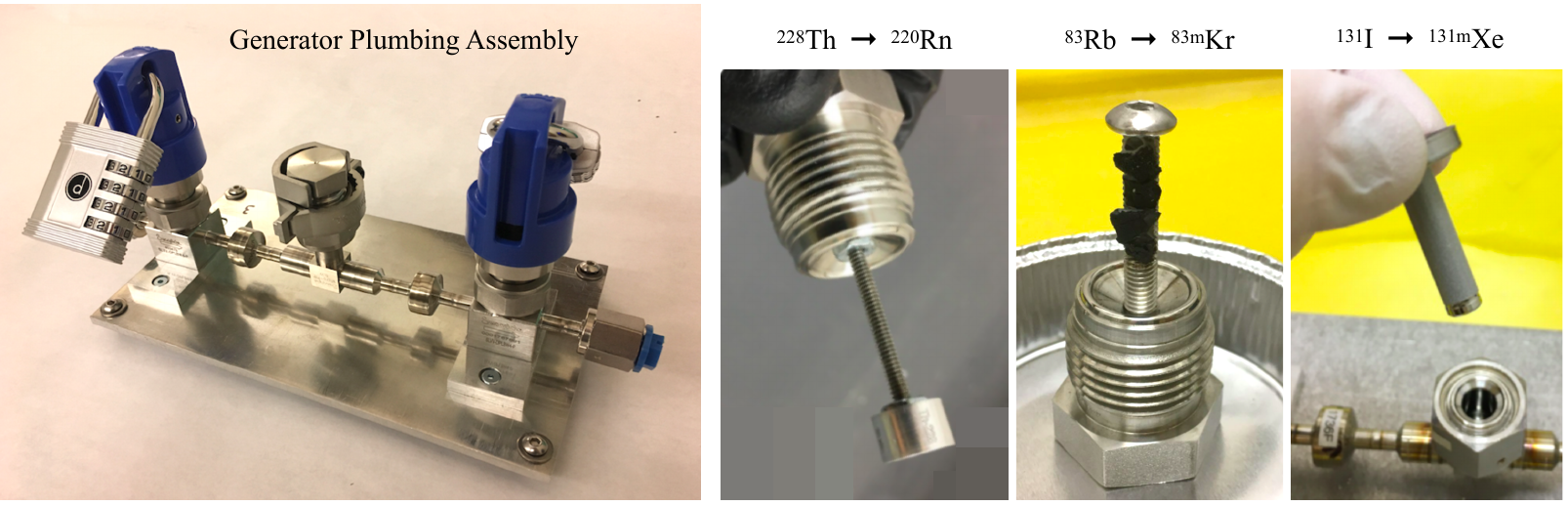}
\caption{An example generator plumbing assembly along with three sets of parent material.  The plumbing assembly is standardized such that any assembly can be mounted in any generator bay on the source injection panel shown in Figure~\ref{fig:injection_panel_P&ID}.  The assembly is welded as one piece to the extent possible, including the 3~nm pore-size sinter filters on either side of the 1/2-inch VCR access port.  Various active parent materials can be installed via this access port:  a platinum disk with electroplated $^{228}$Th for $^{220}$Rn generation, several charcoal pieces dosed with $^{83}$Rb for $^{83\mathrm{m}}$Kr generation, and a VCR gasket combined with a sintered ``cup'' which retains $^{131}$I-dosed NaH$_2$PO$_4$ grains for $^{131\mathrm{m}}$Xe generation.}
\label{fig:generator-assembly}
\end{figure}

An example generator plumbing assembly is shown in Figure~\ref{fig:generator-assembly}.  Each assembly can be interchanged with another assembly, simplifying planning and operations. The parent material is housed within a central 1/2-inch stainless steel (Grade 316) tee, accessible via a 1/2-inch metal face-seal (VCR) cap. This central volume is bounded on either side by sintered-nickel filters (3~nm pore size)~\cite{EntegrisWG3NSMJJ2} to mitigate any granular transport of parent material.  The entire assembly is bounded by two manual locking diaphragm valves~\cite{Swagelok6LVV-DPLBW4-P} which allow mounting to the injection panel via 1/4-inch VCR seals.  The 1/2-inch access port is typically locked in place to prevent loosening during shipping, and the entire assembly is otherwise welded as a single element to eliminate opportunities for leaks to develop. Each generator source has a different mechanism for storing the parent isotope within the central tee as discussed below.

For the $^{220}$Rn source, the $^{228}$Th parent material is procured from Eckert $\&$ Ziegler~\cite{EZsource}, electroplated onto a platinum disk designed to fit inside the generator's 1/2-inch tee. The platinum disk is suspended on one end of a threaded rod whose other end is screwed into the 1/2-inch VCR cap. The 55.6 second half-life of $^{220}$Rn necessitates continuous, high-rate flow of Xe gas through the generator during a calibration run. This flow enables a sufficient fraction of the $^{220}$Rn to reach the getter before its decay into getter-capturable daughters. Because of its many daughters, $^{220}$Rn could be a useful tool for mapping flow in the TPC. The delayed coincidences between $^{220}$Rn$\rightarrow^{216}$Po and $^{212}$Bi$\rightarrow^{208}$Tl pairs, separated by 145~ms and 3.05~min, respectively, can be used for flow-mapping in the TPC LXe. Additionally, the broad beta decay spectrum of $^{212}$Pb is useful in calibrating ER response.

In the case of $^{83\mathrm{m}}$Kr, high surface-area charcoal~\cite{Calgon} is bound to a stainless steel support using low-outgassing epoxy~\cite{MasterBondEpoxy}. The $^{83}$Rb parent material is procured from NIDC~\cite{NIDC} suspended in an HCl solution, which is deposited onto the charcoal at $\upmu$L volume scales. No $^{83}$Rb loss is observed after an initial high-temperature bake. The branching ratio for $^{83}$Rb decay to $^{83\mathrm{m}}$Kr is 78\%~\cite{Hannen_2011}, but due to its relatively short half-life, much of the daughter $^{83\mathrm{m}}$Kr fails to escape the charcoal substrate before decay. As a result, the maximum outgassed $^{83\mathrm{m}}$Kr activity is typically 10\% of the parent $^{83}$Rb activity.  The $^{83\mathrm{m}}$Kr source is useful as a low energy monoenergetic source (a ``standard candle'') for calibrating position-dependent light and charge signal (S1 and S2) collection efficiencies. It is also useful for the measurement of ``electron lifetime", which characterizes the survival probability of ionization electrons as they drift towards the gas phase to produce the S2 signal~\cite{collaboration2020lux, MTimalsina:2022thesis}. The electron lifetime is a gauge of the liquid xenon purity; a higher electron lifetime is observed in purer xenon~\cite{LZ:2022ufs}. Moreover, $^{83\mathrm{m}}$Kr can be used for understanding LXe flow and probing mixing on a few hours timescale, as discussed in section~\ref{injection}. Lastly, $^{83\mathrm{m}}$Kr can also be utilized to derive the electric field map in the TPC through measuring the ratio of the S1 amplitudes of its two decays (at 9.4~keV and at 32.1~keV)~\cite{akerib2017kr}. 

The $^{131\mathrm{m}}$Xe generator technology is created for LZ, anticipating that the LXe mixing timescale within the TPC may be quite long (hours or days), and that a monoenergetic source with a half-life longer than $^{83\mathrm{m}}$Kr ($t_{1/2}$=1.83~h) may be desired. The parent $^{131}$I material is procured in the form of pre-dosed grains of NaH$_2$PO$_4$ salt, distributed by Cardinal Health for thyroid diagnostic measurements~\cite{CardinalHealth}. The salt grains are removed from their medical gelatin capsule and placed into a sintered metal `cup' bonded to a 1/2-inch VCR gasket~\cite{EntegrisGasketgard}. The sintered cup has an average pore size of 300~nm and can safely hold the granular material within the central tee (the 3~nm filter elements remain in place as well). The $^{131\mathrm{m}}$Xe source serves similar purposes as $^{83\mathrm{m}}$Kr (for understanding signal efficiencies, electron lifetime, and LXe flow), but its long half-life makes it the preferred option when LXe mixing timescales are long (days rather than hours). Though its half-life is relatively long, the $\gamma$ decay energy ($\sim$164 keV) of $^{131\mathrm{m}}$Xe is beyond the WIMP energy region of interest and is consequently easily removed as a background for a science search.
\subsection{Dispersed Source Injection and Dose Control}
\label{injection}
\begin{figure}[ht]
\centering
  \centering
 \includegraphics[width=0.47\textwidth]{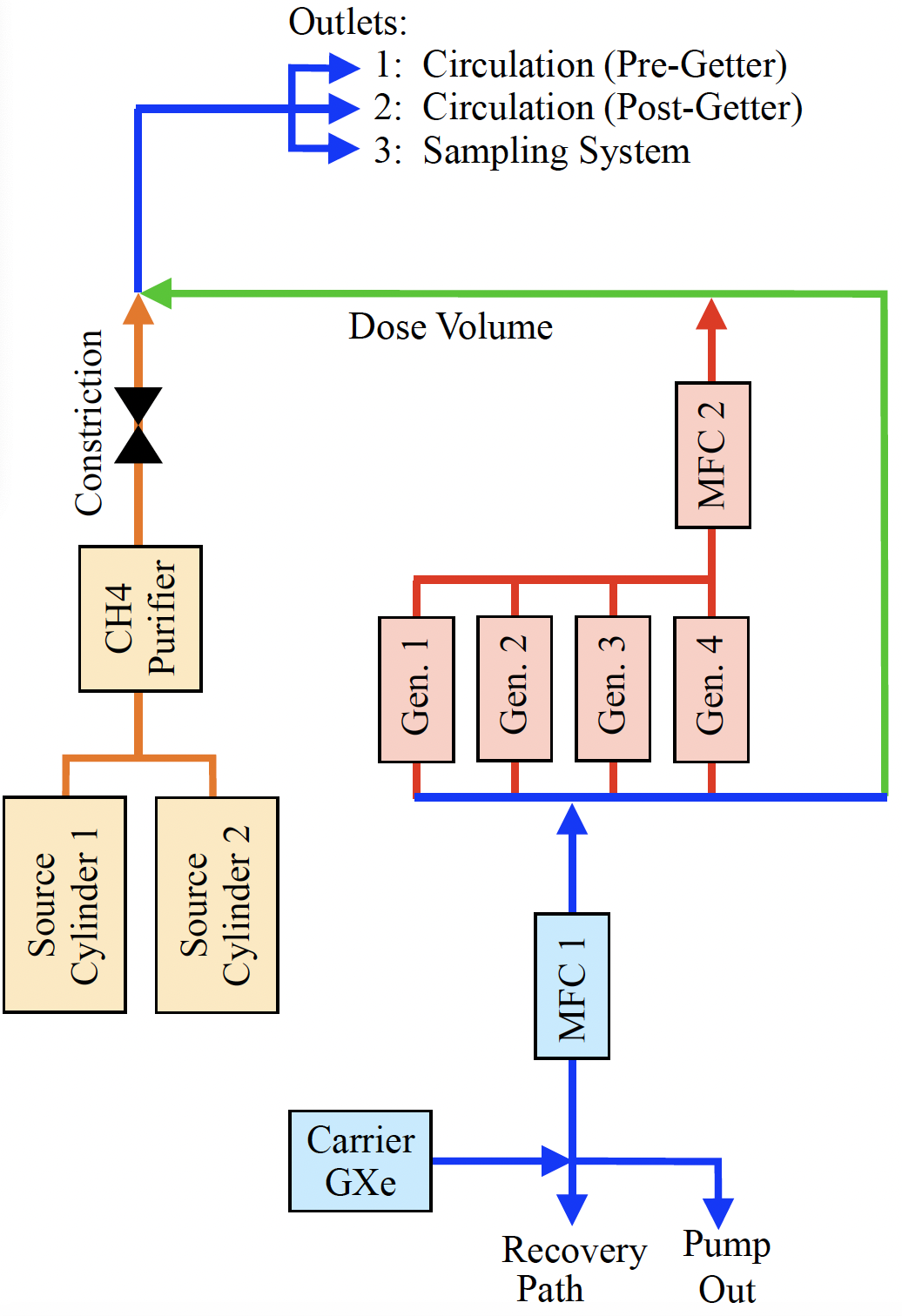}
\caption{Simplified diagram of the source injection panel, highlighting various regions with corresponding colors. The green region is used for dosing $^{83\mathrm{m}}$Kr and CH$_4$-based sources, and serves as a flow through path for the other gaseous sources. Gas flow through the panel is at the scale of a few to several hundred standard cc per minute, set either by a fixed flow restriction or by variable mass flow controllers (MFCs).  Numerous pneumatic valves (not indicated in the schematic) set the source and flow path.}
\label{fig:injection_panel_P&ID}
\end{figure}

The LZ gaseous source injection system, shown in Figure~\ref{fig:injection_panel_P&ID}, is designed to push a precisely controlled quantity of gaseous radioactivity into LZ circulation. It was successfully tested before LZ operations began. In addition to the primary goal of precise dose control, this system also allows for the injection of a wide range of fractions of either parent activity (e.g., $^{131}$I or $^{83}$Rb activity) or stored bottle activity in the case of CH$_4$-based sources.  The panel uses Xe as a carrier gas to transport the injected activity. The carrier gas cylinder is re-filled as needed by cryopumping Xe directly from the main circulation path at a low flow rate, controlled through a dedicated cryopumping mass flow controller (MFC).

\begin{figure}[ht]
\centering
  \centering
  \includegraphics[width=\textwidth]{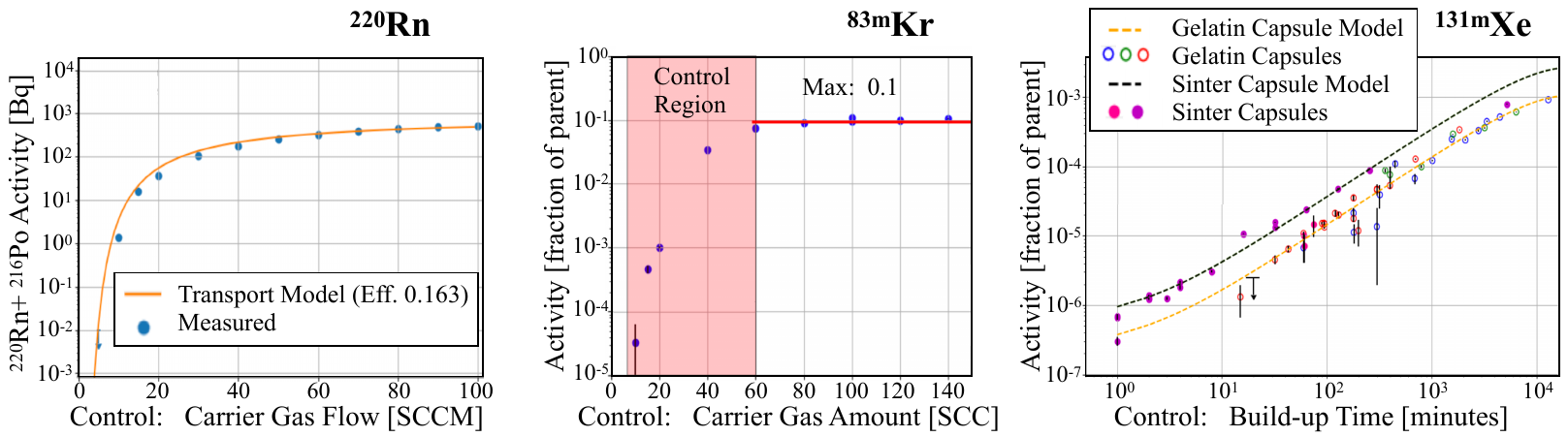}
\caption{Results of dosing control studies performed on a system test version of the source injection panel. In each panel, the x-axis represents the `control parameter' used during injection to set the injected activity. A $^{220}$Rn injection proceeds by a continuous flow of the Xe carrier gas through the generator, with the carrier gas flow rate setting the $^{220}$Rn fraction which escapes the injection panel and reaches the getter before decaying. The $^{220}$Rn figure shows the equilibrium activity increases exponentially with the carrier gas flow rate, which can be described by a transport model from reference~\cite{Nedlik:2022fdx}. This allows us to control the total number of $^{220}$Rn events by adjusting the duration of the injection. For $^{131\mathrm{m}}$Xe and $^{83\mathrm{m}}$Kr, the injected activity is measured as a fraction of the parent material's activity at the time of the injection. In both cases, there is a maximum-possible activity to inject: approximately 10\% for $^{83\mathrm{m}}$Kr and 0.1\% for $^{131\mathrm{m}}$Xe (a product of the branching fraction and the gas emanation efficiency).   The $^{83\mathrm{m}}$Kr injected activity is controlled by pushing only a small fraction of the generator activity out of the generator assembly, with carrier gas flow quantity measured in standard cc.  The comparatively long half-life of $^{131\mathrm{m}}$Xe (and the shorter half-life of its parent $^{131}$I) motivates a dose control strategy based on pumping out the generator contents and then waiting a build-up time specific to the desired $^{131\mathrm{m}}$Xe activity.}
\label{fig:injection_results}
\end{figure}

\begin{figure}[ht]
\centering
  \centering
  \includegraphics[width=\textwidth]{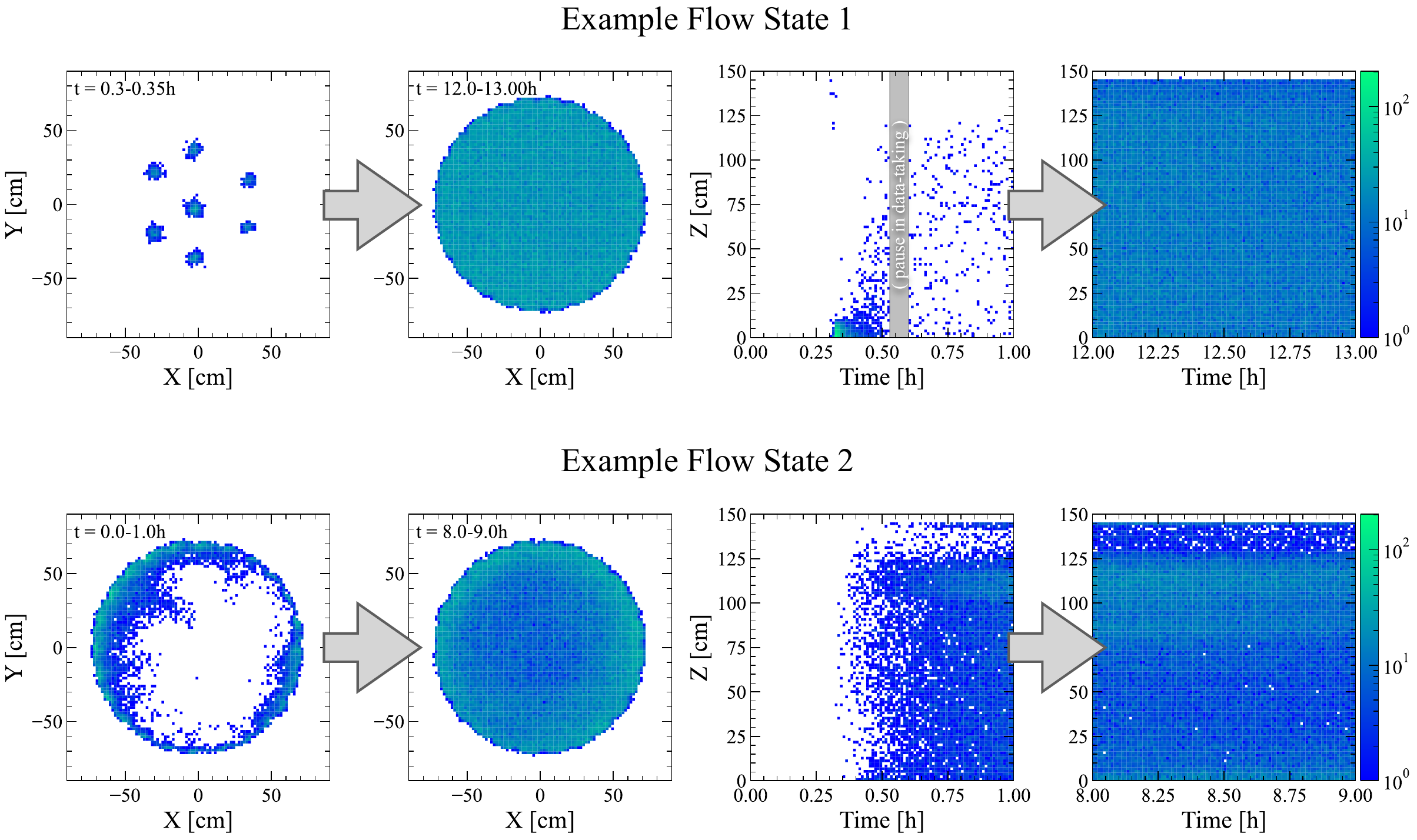}
\caption{Examples of studying LXe flow in the TPC using injected $^{83\mathrm{m}}$Kr. 
Example Flow State 1 shows the first $^{83\mathrm{m}}$Kr injection taken in LZ, which required tuning of the data acquisition (DAQ) settings to optimize data rates. The grey band represents a gap in data taking to change the DAQ settings. The second flow state ``Flow state 2'' is purposefully optimized to increase the calibration source convective mixing efficiency.}
\label{fig:mixing}
\end{figure}

The panel is designed around several isotope-specific methods of dose control.  A variety of control methods are required due to the variety of calibration isotope half-lives. The $^{220}$Rn isotope, with the shortest half-life, is injected using a continuous fast flow through its generator assembly to enable a sufficient fraction to pass the getter before decaying, on its way from the source injection panel to the main xenon circulation system and into the TPC.  In this case, the rate of this constant flow (typically of $\mathcal{O}(100)$~sccm scale) sets an average flow time from the generator to the circulation, and thereby the fraction of the $^{220}$Rn which survives to reach circulation. As shown in Figure~\ref{fig:injection_results} (left panel), this method can robustly control the activity leaving the panel over more than four orders of magnitude. The typical $^{220}$Rn injection time in LZ is about three hours, resulting in a maximum $^{220}$Rn activity of $\sim$50~Hz. Once the injection is stopped, $^{220}$Rn decays quickly, leaving behind $\sim$30 Hz of $^{212}$Pb which is used for our ER calibration. 

The isotope with the next-shortest half-life is $^{83\mathrm{m}}$Kr, which exhibits a strong specific activity gradient along the length of the few-cm generator plumbing.  This gradient results from the $^{83\mathrm{m}}$Kr diffusion distance being cm-scale over the timescale of $^{83\mathrm{m}}$Kr decay. This activity gradient within the generator plumbing greatly aids the dosing control, thanks to the exponential tail of the (roughly Gaussian) specific activity distribution.  A small and carefully metered flow of the Xe carrier gas can push out from the generator a specific portion of that exponential activity distribution. As a result, a linear control of the carrier gas flow results in an approximate logarithmic control of the $^{83\mathrm{m}}$Kr dose. This is accomplished using a low-flow MFC which can control the carrier gas flow at the level of several standard cc.  As seen in Figure~\ref{fig:injection_results} (central panel) showing measurements from a system test, this procedure allows injection control over 3$-$4 orders of magnitude. Typical injections of $^{83\mathrm{m}}$Kr contain activities between 100-200~Bq, which strikes a balance between the number of decays needed for analysis and limitations on activity from event pileup. The obtained event rate is sufficient to study LXe flow and probing mixing as demonstrated in Figure~\ref{fig:mixing}, which depicts two examples of flow states using $^{83\mathrm{m}}$Kr.  In the Example Flow State 1, $^{83\mathrm{m}}$Kr first appears above the cathode directly above each of the seven LXe inlet ports.  Much of this activity disperses below the cathode before greater mixing occurs.  In the example Flow State 2 the $^{83\mathrm{m}}$Kr undergoes significant mixing below the cathode before subsequently crossing the cathode as a diffuse and asymmetric distribution.

The $^{131\mathrm{m}}$Xe isotope has a comparatively long half-life (11.86 days) and its $^{131}$I parent has a shorter 8.03~day half-life.  This short half-life of the parent allows precise dose control by first cryopumping out any gaseous generator contents, and then waiting some specific time (from minutes to days) for the desired $^{131\mathrm{m}}$Xe activity to build up within the generator.  After this specific build-up-time, the gaseous contents of the generator are flushed without requiring precise control. This method allows control over at least 3 orders of magnitude. Two different capsules were tested to store $^{131\mathrm{m}}$Xe in the source generator. The medical gelatin capsule containing the $^{131}$I-dosed salt into the VCR head of the generator is labeled as ``Gelatin Capsules'' in Figure~\ref{fig:injection_results} (right panel). Because of the multi-bar range of pressures experienced by the sources in the injection panel, the gelatin capsule broke within the generator during the test. The sinter cup was thus adopted in order to contain the $^{131}$I powder. This data is labeled as ``Sinter Capsules'' in the same figure. The characteristic activity range of $^{131\mathrm{m}}$Xe which is injected in the TPC to yield useful statistics for the ER calibration is $\sim 1-2$~Bq .

For the bottle sources, the CH$_4$-based sources ($^3$H-labeled CH$_4$ or CH$_3$T, and $^{14}$C-labeled CH$_4$) are mixed with roughly 1 bara (absolute pressure) of Xe carrier gas and stored in cylinders. Their dosing control is accomplished by bleeding out a small fraction of this stored gas mixture until the pressure in a `dose volume' portion of the panel (highlighted in green in Figure~\ref{fig:injection_panel_P&ID}) reaches the desired pressure.  This slow flow is accomplished using a dedicated small-diameter constriction (2~sccm at 35~psid differential), and the dose volume pressure is measured using a high-precision capacitance manometer.  After the dose volume achieves its goal pressure, the contents of the dose volume are flushed into circulation.  

\begin{figure}[ht]
    \centering
    \includegraphics[width=1.01\linewidth]{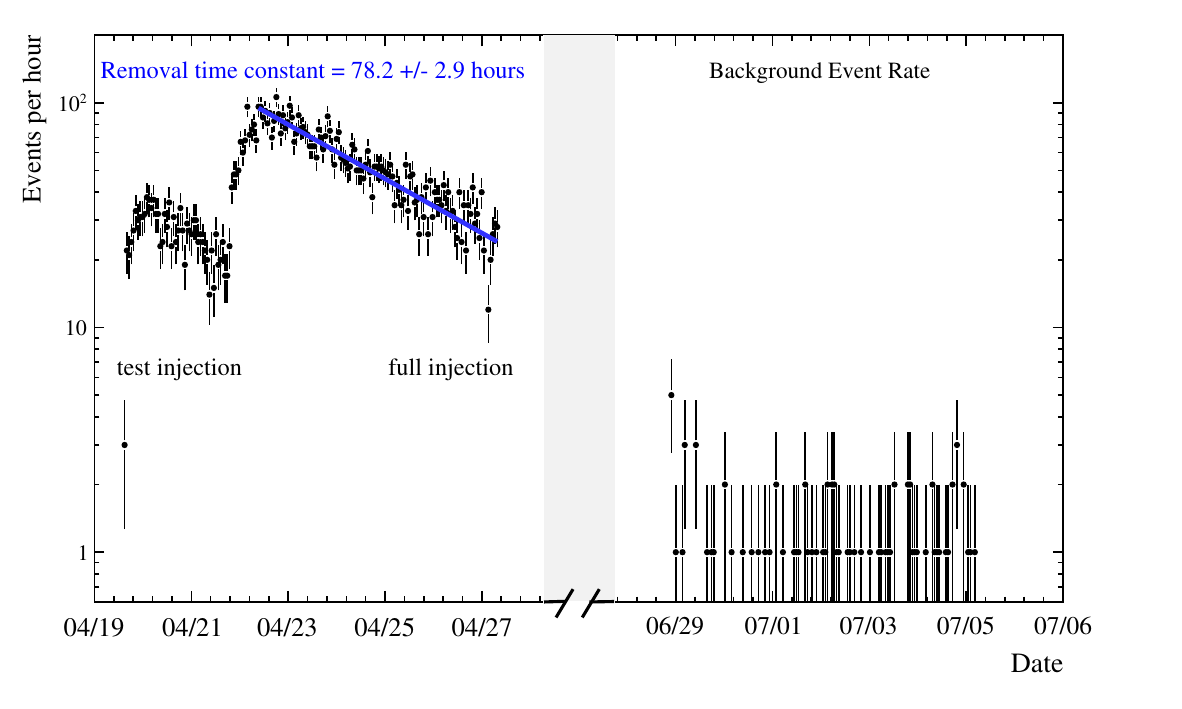}
    \caption{A low-activity test injection of tritiated methane (CH$_3$T) followed by the main injection for an ER calibration in 2022. The removal time constant for CH$_3$T is 78.2 $\pm$ 2.9 hours, measured by fitting an exponential to the time profile of the measured CH$_3$T event rate in the TPC. The exponential fit to the data is represented by the blue line in the plot. The right half shows the normal background event rate during the science run when all the tritium has been removed. The grey band represents the time gap between the tritium calibration and the science run. 
    }\label{fig:CHST_injection}
\end{figure}

The precision dose control is important to mitigate against risks associated with injecting excessive radioactivity, especially long-lived isotopes like CH$_4$-based sources, into the TPC. The system described above enables us to carefully proceed with a staged approach for these sources, starting with small test dose injections before the main calibration in LZ. Successful control of the injected activity was demonstrated at the end of SR1 (which lasted between 23 Dec.~2021 and 11 May~2022). A staged injection approach was used to perform a tritiated methane (CH$_3$T) calibration, with a small test injection followed by a main injection; this is shown in Figure~\ref{fig:CHST_injection}.

Unique among the dispersed sources, the CH$_4$-based sources must be actively removed from circulation due to their long half-lives of many years (the gaseous activity of the three generator sources simply decays in place within the detector). This removal is accomplished using the heated zirconium getter mentioned in Section~\ref{circulation}. The CH$_3$T removal time constant ($\tau$) in LZ has been observed to range from $\sim$65 -- 85~hours depending on circulation rate, getter temperature, and the LXe mixing state within the central TPC. This removal time constant is consistent with the value obtained from pure methane removal using the xenon sampling system during the commissioning phase. An important aspect of the CH$_4$-based source injection hardware is the use of a CH$_4$ purifier~\cite{SAESMC1} within the injection panel.  This purifier removes gaseous species (including free hydrogen and hydrocarbons heavier than CH$_4$) which can contain the long-lived calibration isotope and which can linger for long times on cold detector surfaces.  The use of this purifier stage is therefore essential to the safe use of CH$_4$-based sources in a low-background experiment. For tritium injections in particular, CH$_3$T is used instead of molecular tritium (T$_2$). Due to the larger molecular size and lower diffusion constant and solubility~\cite{10.1116}, CH$_3$T is easier to remove and less likely to diffuse into plastic detector components, which would contaminate the LXe during the WIMP search run. The removal time constant of CH$_3$T from xenon circulating through the getter at a flow rate of 380~slpm was measured in the first science run to be 78.2 $\pm$ 2.9 hours, as shown in Figure~\ref{fig:CHST_injection}. From the same figure, it can also be observed that the background rate post-injection in July is consistent with the background rate from a benchmark run pre-injection in mid-April. 

\section{External Rod Source Calibration} \label{csd}

\subsection{Design and Implementation}\label{CSDintro} 
The external Calibration Source Deployment (CSD) system in LZ lowers neutron and gamma rod sources in three tubes that are positioned in the vacuum space between the inner and the outer cryostat vessels. Each tube is connected to an independent deployment system described below. As a source is deployed at different depths inside the calibration tube, it can generate signals in various regions in the OD, TPC, and Skin detectors. This provides calibrations of the detectors' energy scale (i.e., the observed PMT light signals to energy depositions in these detectors) and the spatial dependence of that energy scale, as well as the inter-detector timing measurements between the OD, the Skin, and the TPC. These timing measurements are critical for applying veto selections that are based on timing to remove background events.

\subsubsection{Design}\label{design_Sec}
\begin{figure}[ht]
 \centering
 \includegraphics[width=0.99\textwidth]{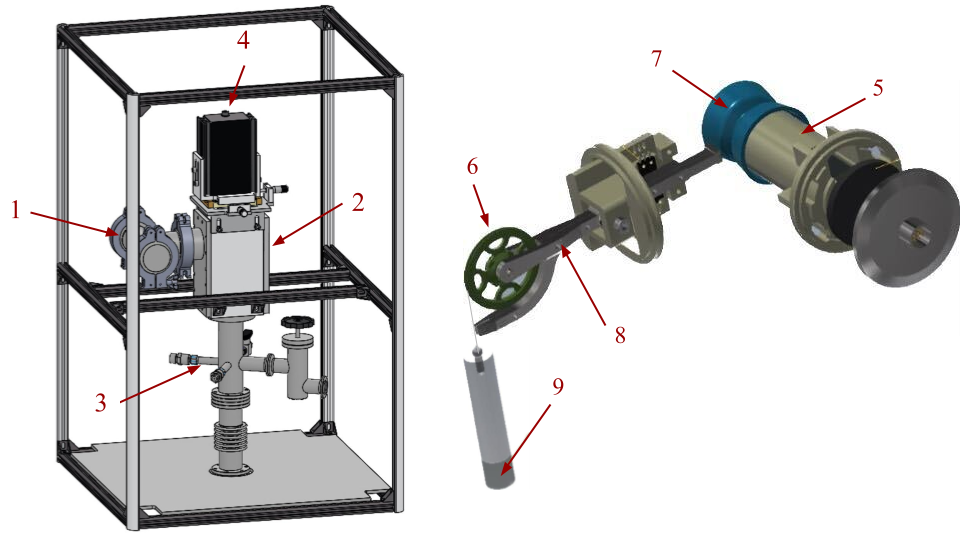}
 \caption{Left: CAD drawing of the main components of the CSD system caged in its protective frame: (1) Tee-Piece containing the deployment mechanics; (2) Source connection chamber with deployment wheel and filament lever; (3) 6-way connection piece with N$_2$ purge, pressure relief, and vacuum valves; (4) Laser position feedback system. Right: the internal deployment system concealed within the T-piece (1) including a stepper motor (5), a deployment wheel (6), a filament drum (7), a lever (8) and a calibration source (9).} \label{fig:csd-head}
\end{figure}

There are three CSD units spaced 120$^\circ$ apart, to provide detector response calibrations at different azimuthal angles. Each CSD unit has two main components: a bottom portion made of a $\sim$ 6~m long rigid stainless steel CSD tube located $\sim$12~cm from the TPC wall (in the inter-space between the inner and outer cryostat vessels), and a top portion containing the ``CSD head'' which controls the operation of the system. This top portion is caged in protective frames seated on the deck above the water tank, and it is coupled to the bottom tube through a 6-way connection piece and bellows, as shown in Figure \ref{fig:csd-head}. The 6-way connector provides a connection to a vacuum port, a pressure gauge to read the vacuum level inside the tube, a N$_2$ line used for purging air inside the CSD tube to mitigate against airborne radon, and a pressure relief valve. The bellows allow slight mechanical adjustment of the system during its installation and help mitigate thermal contraction of the calibration tube when it is cooled down to cryogenic temperatures. 

Each CSD head independently controls a CSD system during source deployment. As such, all three units can be operated simultaneously to deploy the sources to various $z$-positions in calibration tubes as required. Calibration sources need to be deployed to a precision of $\pm$ 5~mm, 
and the system needs to deploy sources to the same $z$-position repeatedly. This has been achieved by the intricate design of the CSD head shown in Figure~\ref{fig:csd-head} (left). The CSD head consists of a 3D-printed source deployment wheel, a lever which guides a $\sim$6~m long filament inside the tube, a high gear-ratio stepper motor~\cite{maxongear} that controls the winding and unwinding of the filament via the filament drum connected to its shaft, and a ferromagnetic source holder connected to the filament that screws onto the calibration sources. The CSD heads are further equipped with end-switches which are activated when there is a sudden change in the filament tension. This typically happens when the rod source reaches either the very bottom of its deployment range or the very top of the deployment range. Upon an end-switch trigger, the system is programmed to reverse direction by 10,000 motor steps (corresponding to $\sim$ 4~cm) and then abort further movement. The end-switch can also be triggered by irregular unspooling of the filament during deployment, providing a safety measure to prevent the system from operating beyond its physical limits.

The temperature of the stepper motors is monitored by thermometers that are permanently fixed to the motors. During operation of the CSD units, it is important to mitigate any temperature rises that might compromise the 3D printed components. In addition to the thermometry and a software protocol that switches off the power to the system in case of overheating, each CSD unit has a temperature switch incorporated in the stepper motor electrical circuit that physically disconnects the power to a specific stepper motor above a preset temperature of 100~$^{\circ}$C.

At the top of the CSD head is a laser light monitoring system that reads out the depth of the source inside the calibration tube. The laser light is reflected from a photo-reflective surface of the source holder, providing optical position feedback. The laser system ensures that the position of the source is determined by an in-situ measurement, providing a primary and separate determination of source coordinates in addition to using the stepper motor.

A major concern one needs to mitigate against is the breaking of the filament holding the source. Extensive tests have been performed on various types of deployment filaments to ensure that they can suspend the weight of a source within a defined safety margin. The maximum weight allowed for each source is $\sim$150~g. Stress tests and fatigue tests have been conducted on filaments, to determine characteristics such as elasticity and tensile strength. In addition, a spring-shaped 3D printed shock absorber is placed at the bottom of each calibration tube to absorb the impact in case a source detaches from the filament and falls to the bottom of the CSD tube. Furthermore, the calibration source holder on the CSD head is made from a ferro-magnetic material which is thread locked to the source during deployment, allowing easy retrieval of the source with a magnet whenever needed.

\subsubsection{Implementation}
\begin{figure}[ht]
 \centering
 \includegraphics[width=0.8\textwidth]{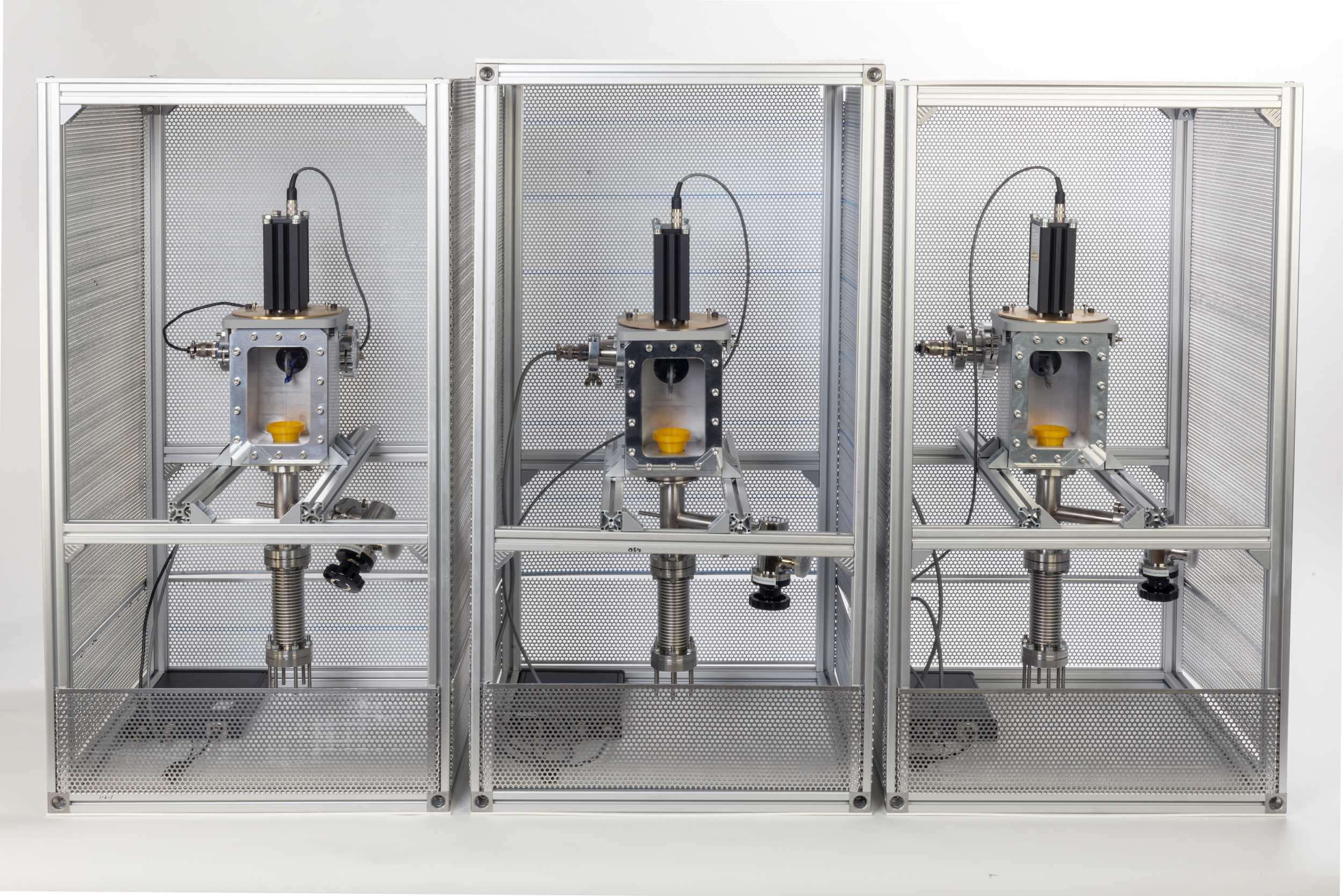}
 \caption{The three CSD units fully assembled before their installation at SURF.}\label{fig:CSD3units}
\end{figure}

The on-site installation of these units involves various steps to optimize their performance. Each of the three CSD units is built up in its support frame and coupled to the connecting flange at the top of the calibration tube. The frame contains support bars that are adjustable in the $x$-$y$ plane and in height ($z$), and a laser adjustment plate controls the angle at which laser light enters the calibration tube. A separate CSD electronics unit houses the field-programmable gate array (FPGA) board that enables remote computer control of the source deployment. It provides power to the CSD motors, the data transfer between the CSD units, and the LZ slow control system~\cite{LZ:2019sgr}. Figure~\ref{fig:CSD3units} shows a picture of the three CSD units taken before their installation on-site. A check is performed post-installation of each CSD unit by taking a position reading with no source connected to the system. This returns the expected distance from the laser sensor to the shock absorber that sits at the bottom of the calibration tube. Non-radioactive dummy sources and actual sources are also deployed at various depths to ensure mm accuracy in recorded positions.

\begin{figure}[ht]
\centering
  \includegraphics[width=0.89\textwidth]{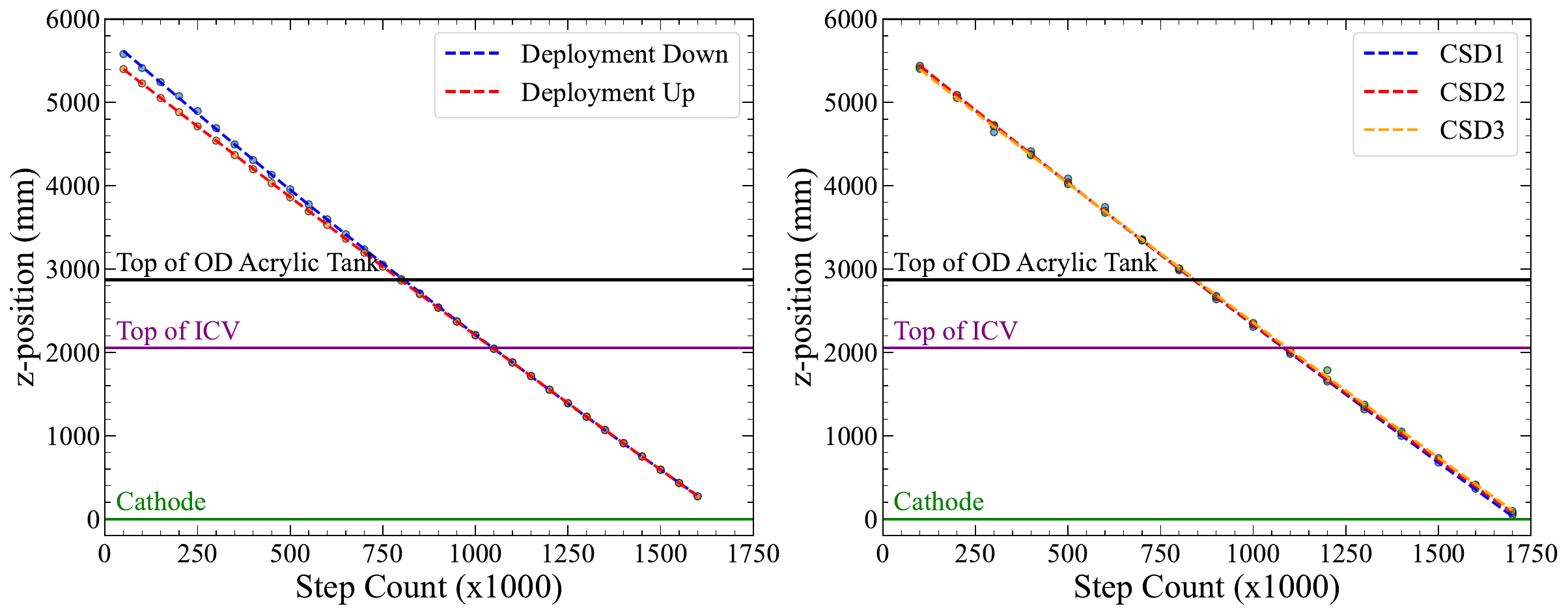}
\caption{Left: Relationship between the physical measurement of steps from the stepper motor and the laser's $z$-position reading for a deployment of a non-radioactive dummy source down and up the calibration tube. A discrepancy between the two deployments is observed when the source is first installed, due to the filament adjusting to the new weight. Right: Deployment alignment curves for three CSDs when the filament is fully stretched, showing achieved deployment positions for a full range of desired heights for calibrations. The positions of the TPC cathode, ICV and top OD acrylic tanks are marked on both plots. } 
\label{fig:csd-align-plots}

\end{figure}

During the source deployment, the CSD relies on optical position feedback of the laser to achieve a $\sim$mm position accuracy. A crucial element for optimal performance is the precise alignment of the laser with the CSD tube, as any misalignment from the laser could cause a misreading of the source position. We established a ``laser alignment calibration protocol'' which consists of producing a series of pre-calibrated "deployment curves" for each CSD unit. Each curve represents the deployment distance measured versus the number of steps the motor is driven and confirms the alignment of the laser. An approximately linear relationship is obtained from these deployment curves as shown in Figure~\ref{fig:csd-align-plots}, with a slight discrepancy noted when a newly-installed source is deployed downward and upward for the first time. This is indicative of the changing tensile properties of the filament during the deployment process. Upon characterizing this deployment alignment curve, any problematic regions or errant readings are able to be easily identified, and the laser is finely re-positioned to rectify erratic readings. The $\sim$ 0~mm position in the LZ coordinate system is at the cathode, and the highest position ($\sim$ 6,000~mm) is about a meter above the water tank, ensuring a full coverage of the TPC and top OD acrylic tanks. The bottom of the CSD tube is $\sim$1000~mm above the bottom of the side OD acrylic tanks, not allowing their full coverage. Future experiments could benefit from an extended source tube to cover the entire range of all detectors, including the OD.

\begin{figure}[ht]
 \centering
 \includegraphics[width=1.02\textwidth]
 {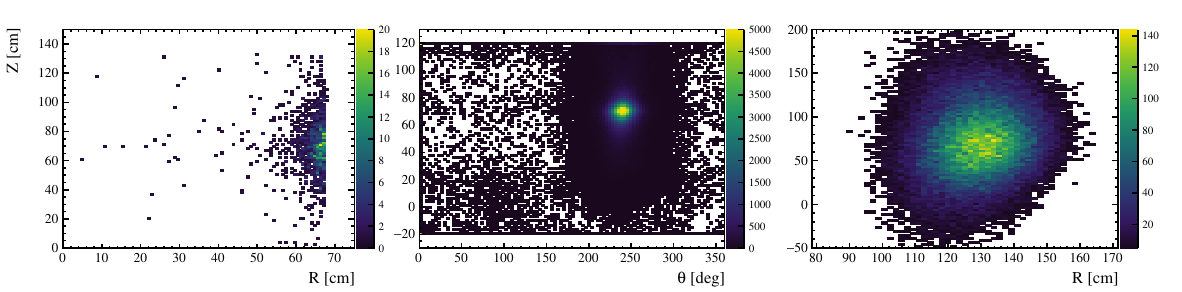}
 \caption{Reconstructed positions of events in the TPC (left), Skin (middle) and OD (right) induced by a $^{228}$Th source deployed to a position of 70~cm above the TPC cathode level in the CSD tube. The x- and y-axes have different scales for the three plots to accommodate the varying dimensions of the three detectors. The x-axis of the middle plot represents the angle ($\theta$), not the radius (R), because the Skin's width in the radial direction is much narrower compared to that of the TPC and the OD.} \label{fig:CSDZvsR}
\end{figure}

Figure~\ref{fig:CSDZvsR} illustrates the detector response to a $^{228}$Th source deployed by the CSD to a position of 70~cm above the cathode. The reconstructed $z$-position of its response in the TPC peaks at ${z = 70.1 \pm}$ 0.7~cm and demonstrates a successful source deployment to the target location. 

\subsection{CSD Neutron and Gamma Rod Sources}

\subsubsection{AmLi Sources}\label{sec:AmLi}
To calibrate the TPC nuclear recoil response and the neutron tagging efficiency of the OD and the Skin veto system, LZ utilizes several neutron sources. AmLi sources made of a mixture of ~$^{241}$Am $\alpha$-radioactivity and $^{7}$Li produce low-energy neutrons emission  via nuclear $(\alpha, n)$ reactions. These sources are used for low energy NR calibration, since their maximum neutron energy of $\sim$1.5~MeV results in nuclear recoils depositing up to $\sim$45~keV$_{nr}$ of energy, close to the maximum nuclear recoil energy in the standard WIMP analysis~\cite{LZ:2022ufs}.

AmLi sources with sufficiently low neutron emission rate (that can avoid pile-up of overlapping events in the LZ detector) are not commercially available and therefore are custom-made by the LZ collaboration. Three such sources are developed, assembled and characterized. The making of the sources and extensive tests to meet the stringent design goals are described in detail in reference \cite{Sazzad:2023uqs} and are briefly describe here. The $^{241}$Am is obtained from Eckert $\&$ Ziegler \cite{EZsource} in the form of Am(NO$_3$)$_3$ in nitric acid solution. After adjusting the concentration of the procured solution, it is centrally deposited on a 0.1~mm gold foil by drops of 10 $\upmu$L. The gold foil holder is then placed on a hotplate inside a fume hood and heated to evaporate the nitric acid solvent. The gold foil with the deposited $^{241}$Am is then wrapped in a 0.75~mm lithium foil to sandwich the $^{241}$Am. This process is carried out inside a glove box filled with argon to prevent the lithium foil from oxidizing. The wrapped foil mixture is then encapsulated in three nested metal capsules: two inner epoxy-sealed tungsten shielding capsules intended for suppressing the gamma-radiation emitted following $^{241}$Am $\alpha$-decays, and an outer stainless-steel cylinder cover designed to integrate in the CSD system. The encapsulated sources are then leak tested by soaking them into nitric acid per the ISO-9978 standard~\cite{ISO9978}. Any leakage of the americium into the acid would result in a distinct 59~keV peak in a gamma screening detector. Results of the screening of the soak acid from all three sources show no evidence of leakage. A limit of $<$ 5.0~mBq at 90$\%$ confidence level is placed on the $^{241}$Am activity leaked out of the source capsules, corresponding to a fractional limit of ${< 5.3 \times 10^{-11}}$.

\begin{figure}[ht]
 \centering
 \includegraphics[width=0.85\textwidth]{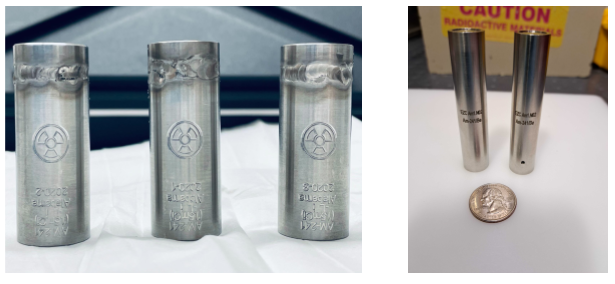}
 \caption{Left: Three custom-made AmLi sources used in LZ. The gold-americium-lithium matrix is encapsulated in nested tungsten and stainless steel capsules, which are designed to integrate with the CSD system and provide gamma-radiation suppression. The welds on the outer stainless-steel capsules are visible in the picture. Right: Two customized AmBe sources from Eckert $\&$ Ziegler designed for LZ. The stainless steel threaded capsules were designed to integrate in the CSD system and shield against low energy gammas from $^{241}$Am decay. } \label{fig:AmLi}
\end{figure}

\begin{figure}[ht]
 \centering
 \includegraphics[width=0.8\textwidth]{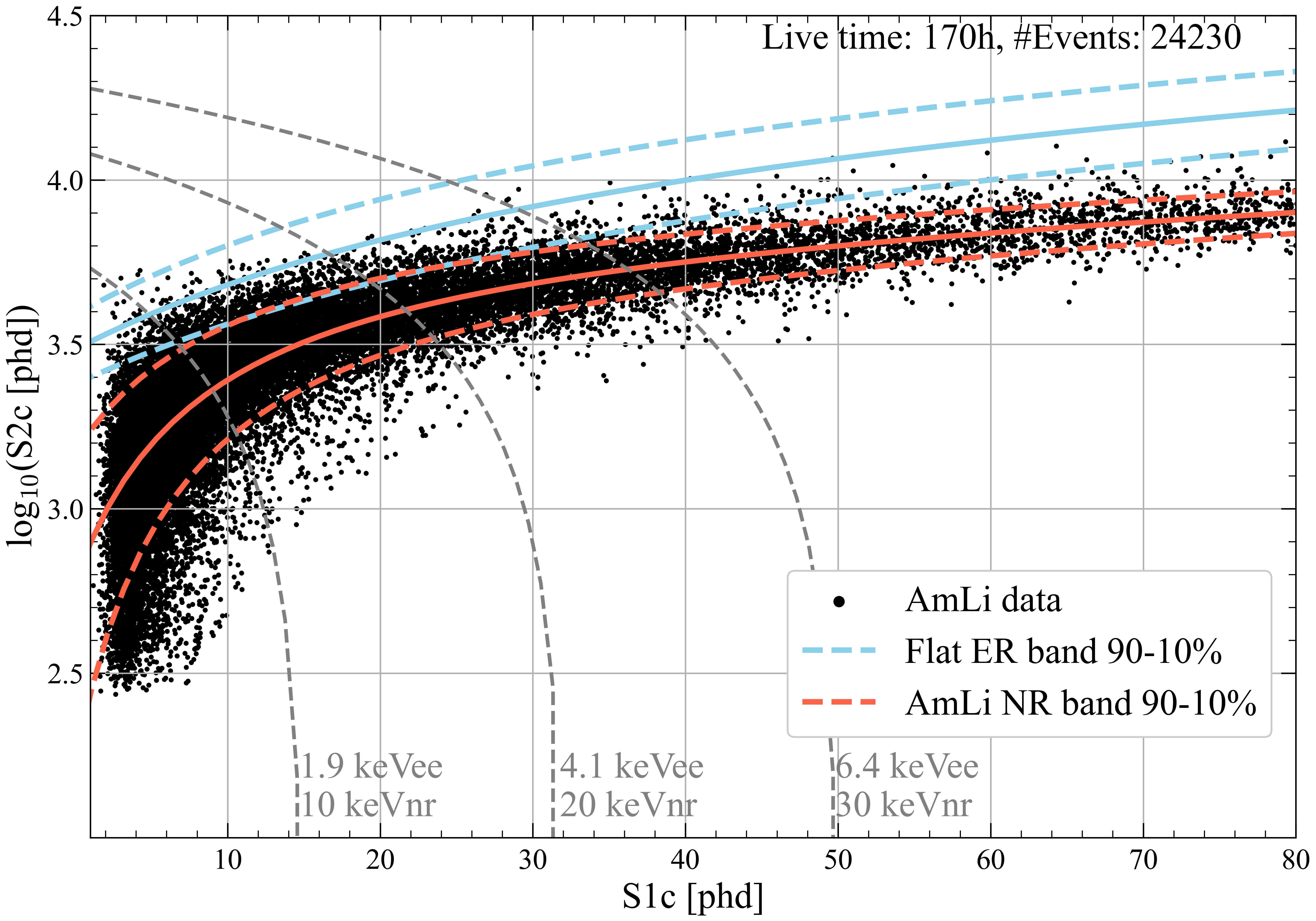}
 \caption{Detector response to the AmLi source in LZ. The solid (dashed) blue line is the median (90-10\% quantiles) of a flat ER distribution modelled using $\textsc{nest}$ 2.3.7~\cite{szydagis2011nest}. The solid (dashed) red line is the median (90-10\% quantiles) of an NR band fitted to the AmLi data using a skewed Gaussian distribution~\cite{PhysRevD.102.112002}. The dashed grey lines are contours of constant energies. Only data points within 5-sigma of the fitted AmLi distribution are shown.}\label{fig:AmLidata}
\end{figure}

After the successful leak testing of the sources, the sources are calibrated for their neutron and gamma emission rates before being shipped to SURF. The neutron emission rates are determined with a setup consisting of four RS-P4-0813 $^{3}$He proportional tube counters~\cite{Reuter-Stokes} suspended in a water tank, with water used as moderator for fast neutrons. The emitted gamma-spectra are carefully studied using measurement data from a high-purity germanium detector and thorough Monte Carlo simulations. The obtained $^{241}$Am activity, gamma emission rates, neutron emission rates, gamma-to-neutron ratio, and neutron yields of the three sources are tabulated in Table~\ref{tab:tableWithAllInfo}. Figure~\ref{fig:AmLi} (left) shows a picture of the manufactured AmLi sources, and Figure~\ref{fig:AmLidata} shows the S1 and S2 signal distribution induced by AmLi neutrons from a 170 live hour run in LZ. In SR1, these sources not only played an essential role in calibrating the TPC NR response, but also in facilitating measurements of neutron backgrounds by determining the neutron tagging efficiency of the veto systems. Moreover, a combined tritium and AmLi dataset was used to evaluate the signal efficiency that was used in the WIMP-search analysis~\cite{LZ:2022ufs}. The OD neutron tagging efficiency was determined through the inter-detector coincidence induced by a neutron in the TPC and its capture signal in the OD. After subtracting accidental coincidences between the TPC and OD caused by high-energy gammas and neutrons from the calibration source, as well as background radioactivity seen by the OD, the position-averaged single scatter AmLi neutron tagging efficiency was measured to be $89 \pm 3$\%~\cite{LZ:2022ufs} using the Skin and OD combined. However, background neutrons come from ($\alpha, n$) reactions and fission processes in detector materials, and they are thus accompanied by $\gamma$-rays. The AmLi neutrons do not have a correlated prompt gamma emission. This may result in a lower tagging efficiency for AmLi neutrons as compared to background neutrons (i.e., an overestimate of the number of background neutrons). This systematic did not affect the SR1 results, since zero neutrons are observed in the relatively short exposure, but it will be a factor for future science runs with longer exposures. An ($\alpha, n$) source with correlated gammas, such as the americium-beryllium source, which mimics more closely the background neutrons, is therefore considered for improving the calibration of the neutron tagging efficiency.
\begin{table}[htb]
    \begin{center}
    \begin{tabular}{p{1.3cm} p{2.0cm} p{2.2cm} p{2.2cm} p{2.2cm} p{2.3cm}}
    \hline
    Source & $^{241}$Am activity [MBq] & $\gamma$-ray emission rate [Hz] & Neutron emission rate [Hz] & $\gamma$-to-neutron ratio & Neutron yield [n/10\textsuperscript{6} $\alpha]$ \\
    \hline
    AmLi-1 & 31.2 $\pm$ 1.4 & 368 $\pm$ 59 & 18~$\pm$~2 & 20~$\pm$~4 & 0.59~$\pm$~0.06\\

    AmLi-2 & 20.3 $\pm$ 1.0 & 239 $\pm$ 38 & 9~$\pm$~1 & 26~$\pm$~5 & 0.45~$\pm~$0.04\\

    AmLi-3 & 27.0 $\pm$ 1.2 & 318 $\pm$ 51 & 12~$\pm$~1 & 30~$\pm$~6 & 0.46~$\pm$~0.04\\
    \hline
    \hline
    Total & 79.0 $\pm$ 2.0 & 925 $\pm$ 87 & 39~$\pm$~3 & & \\
    \hline
    \end{tabular}
    \end{center}
\caption{Individual AmLi source's $^{241}$Am activity, gamma emission rate, neutron emission rate, gamma-to-neutron emission ratio, and neutron yield~\cite{Sazzad:2023uqs}. }\label{tab:tableWithAllInfo}
\end{table}

\subsubsection{AmBe Sources} 
In addition to AmLi sources, americium-beryllium (AmBe) neutron sources have been planned for deployment in the CSD system. Neutrons emitted from AmBe sources have energies of up to $\sim$11~MeV, which can produce xenon recoils of up to $\sim$330~keV$_{\mathrm{nr}}$ in the TPC,  enabling the calibration of a wider range of NR energy signals. In non-standard WIMP analyses, such as EFT (Effective Field Theory) analyses where WIMP recoil spectra exhibit substantial rates in the energy range well beyond the AmLi source endpoint energy~\cite{aalbers2023first,akerib2021effective}, higher energy neutron sources like AmBe are useful for calibrating the detector's NR response.

The AmBe sources can also be used to cross-check measurements of the NR detection efficiency in the standard WIMP analysis and improve the neutron tagging efficiency measurement in the OD, which is crucial for vetoing neutron backgrounds. The prompt 4.4~MeV gammas, emitted alongside the neutron ${\sim 58\%}$ of the time \cite{liu20074,geiger1975radioactive} through the AmBe reaction $\alpha + ^{9}$Be $\rightarrow ^{13}$C$^{*} \rightarrow ^{12}$C + n + $\gamma$, can provide a powerful signature for tagging the neutron as they provide signals in coincidences among the OD and Skin veto detectors and the TPC. Additionally, as mentioned in section~\ref{sec:AmLi}, high energy gammas from the $^{241}$Am decay can be mistaken for those generated by neutron captures in the OD, resulting in accidental TPC-OD coincidences that have to be corrected to determine the OD tagging efficiency. Hence, an AmBe source producing fewer gamma backgrounds for a comparable neutron rate is important for verifying the measurement results and constrain their uncertainty. Beryllium has the highest neutron yield among the light elements for ($\alpha, n$) reactions~\cite{alphN} and therefore the ${}^{241}$Am($\alpha, n$)$^9$Be has a lower gamma-to-neutron ratio compared to ${}^{241}$Am($\alpha, n$)$^7$Li. In principle, with a lower ${}^{241}$Am activity (e.g 50~$\upmu$Ci), a $\times 3$ neutron yield (e.g 10$^2$~n/s) than the three AmLi sources can be achieved while greatly reducing the gamma rate. The reduced gamma rate is also important for the calibration of the TPC as it can reduce the contamination of neutron events with activity from the associated gammas.

We have obtained a $\sim$ 130 $\upmu$Ci custom-made AmBe source from Eckert $\&$ Ziegler (see Figure~\ref{fig:AmLi}  (right)) and performed preliminary calibrations using that source. In addition, we are exploring other AmBe source designs that have lower activities and/or utilize external tagging of the 4.4~MeV gamma for more precise neutron selection which will optimize the neutron tagging efficiency measurement and the TPC NR calibration in LZ.

\subsubsection{Gamma Rod Sources}
\begin{figure}[ht]
 \centering
 \includegraphics[width=0.5\textwidth]{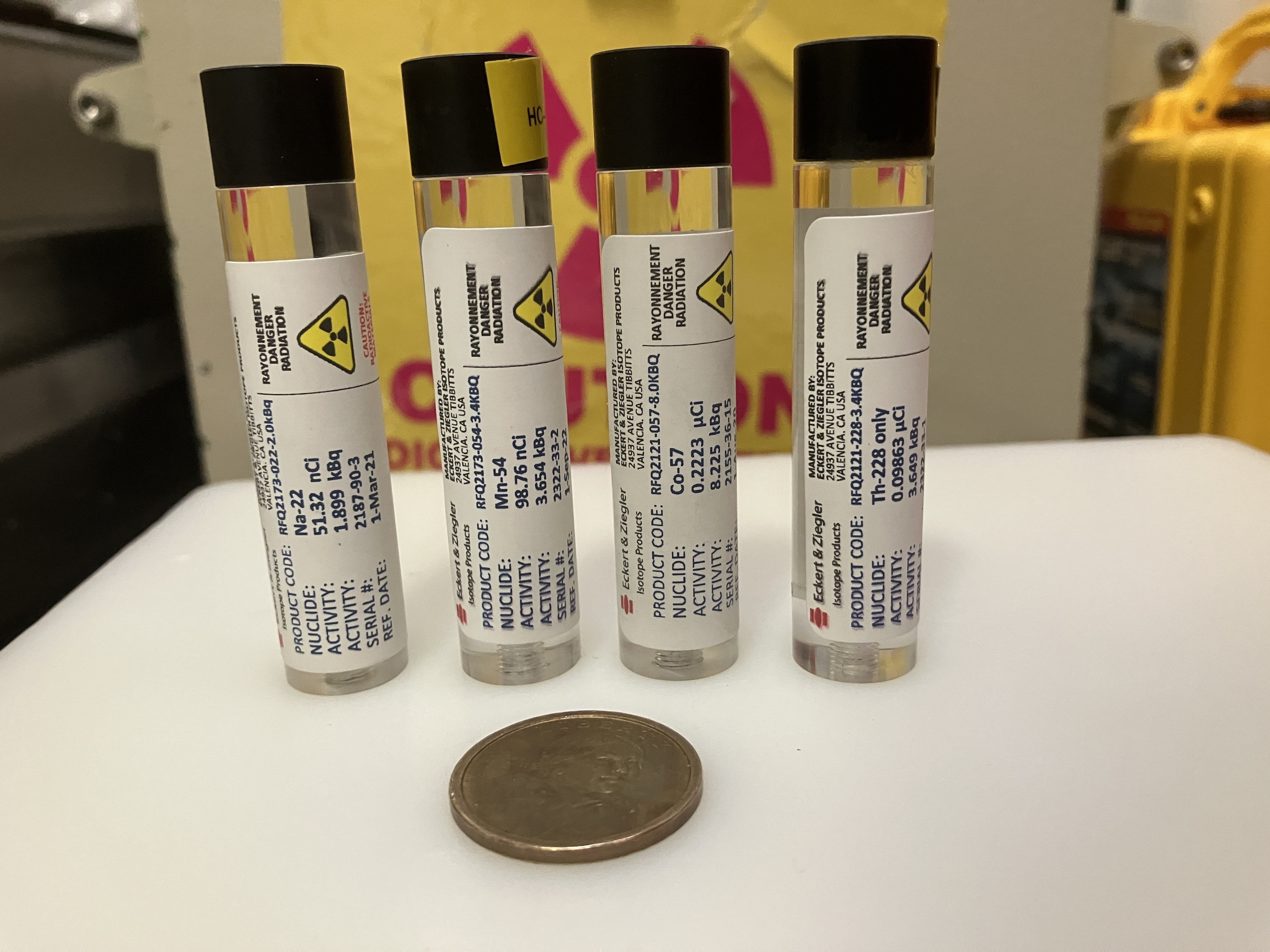}
 \caption{Picture of $^{57}$Co, $^{54}$Mn, $^{22}$Na, and $^{228}$Th $\gamma$-ray sources that are deployed in LZ for high energy ER calibrations.} \label{fig:EZsource}
\end{figure}

The external gamma sources used in LZ are commercial rod sources procured from Eckert \& Ziegler. Because LZ is a low-background experiment with a typical trigger rate of ${\sim 15}$~Hz during science searches~\cite{LZDAQ_paper}, 
the source activities needed for the gamma calibration do not need to be very high ($<150$~Hz in all three detectors). The geometry of the sources is custom-designed in order to be integrated with the CSD system. Each source is made by sealing a small amount of radioactive salts in the tip of a $\sim$50~g, 15.9~mm diameter $\times$ 74.9~mm acrylic rod. The weight of the source is well within the suspension capacity of the CSD and makes it easy to deploy during the calibration. The inactive end of the source has an M4 tapped hole to mate the source holder on the CSD head. Figure~\ref{fig:EZsource} shows four types of sources deployed in LZ that emit $\gamma$-rays of different characteristic energies: $^{57}$Co, $^{54}$Mn, $^{22}$Na, and $^{228}$Th. 

\begin{figure}[ht]
 \centering
 \includegraphics[width=1\textwidth]{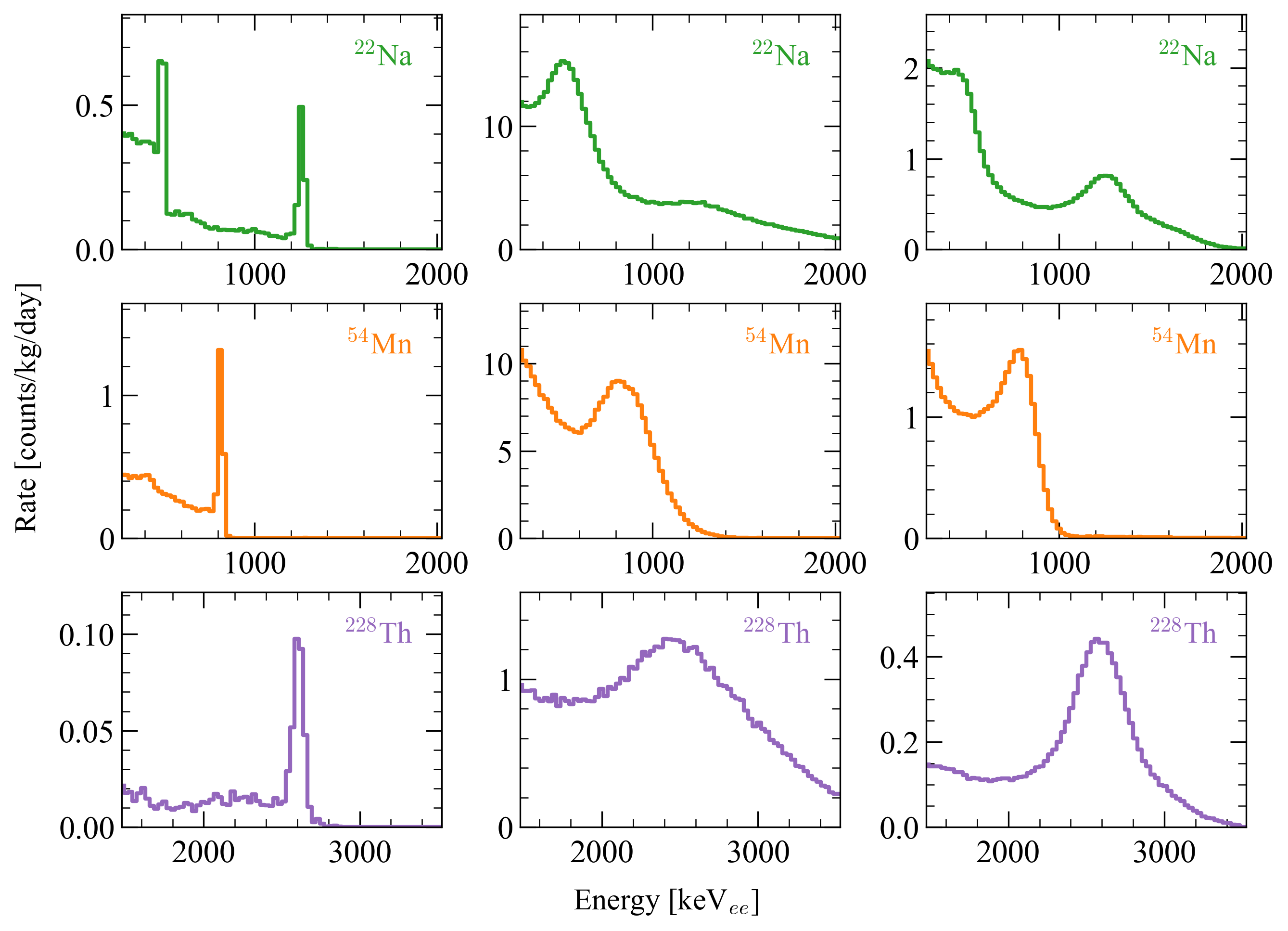} 
 \caption{Measured energy spectra from $^{22}$Na (511~keV from positron-electron annihilation and 1274~keV gammas), $^{54}$Mn (835~keV), and $^{228}$Th (2615~keV) in the TPC (left), Skin (middle) and OD (right). Only single scatter events are plotted. The energy range for the $^{228}$Th data is different from the range used for the $^{22}$Na and $^{54}$Mn data. Data are taken using the above three gamma sources positioned at $z$=70~cm above the TPC cathode. Spectral peaks from these sources are reconstructed at their expected values.
 }
 \label{fig:gammapeaks1d}
 \end{figure}

These sources are used in LZ to characterize the TPC detector response to high energy ER signals such as $0\nu\beta\beta$ of $^{136}$Xe and gamma backgrounds from detector materials or the rock of the experimental cavern. Once the energy scale is well understood from the calibration, we can measure the rates of background components by fitting their mono-energetic peaks at high energies and extrapolate their contributions to the low energy range relevant for the WIMP search~\cite{LZ:2022ysc}. 

The energy spectra in the TPC, Skin, and OD produced by three of these gamma sources, $^{54}$Mn (835~keV), $^{22}$Na (511~keV from positron-electron annihilation and 1274~keV), and $^{228}$Th (primarily for 2615~keV), spanning the $\sim$500-2700~keV$_{ee}$ energy range, are shown in Figure~\ref{fig:gammapeaks1d}.
The mean free path for 122~keV $\gamma$-rays from $^{57}$Co is $<$ 3~mm in LXe~\cite{NISTXCOM} and thus only a small fraction of these gammas can make it into the TPC; they are mostly stopped in the skin region. Therefore, $^{57}$Co is mainly used for the Skin and OD energy calibrations in normal operations. During the commissioning of the LZ detector, the $^{57}$Co source was positioned above the liquid surface and provided calibration data for the TPC liquid leveling, complementing leveling data taken by the weir precision sensors and Skin PMTs. As the attenuation of the $\gamma$-rays from $^{57}$Co is smaller in the gas phase, the leveling of the liquid surface could be calibrated using induced S2s from this source. At a constant gate-anode voltage ${\Delta}$V, the pulse width of an S2 varies inversely with the distance between the liquid level and the anode grid. By adjusting the tilt of the detector and observing the S2 distributions using the $^{57}$Co source, the detector was iteratively leveled using these in-situ measurements.

The Skin and OD energy calibrations are crucial for understanding the response of the veto systems and are carried out by deploying gamma sources at multiple $z$- and $\theta$- positions in the CSD tubes. For the OD, the position scan is necessary as its complex geometry and material components with different optical properties can cause photon detection efficiency to vary greatly by the location of particle interaction. For the Skin, besides its geometry that affects light collection, its large electric field variations (a factor of $>$100 higher at the top and bottom compared to the center, according to simulations) can result in largely different light yield at different locations. The comprehensive calibration using sources of different energies at various locations provide essential information for correcting the location-dependent detector response. Beside high-energy ER calibration, these gamma sources were also used for timing measurements among the three detectors. This inter-detector timing is then used to provide input for vetoing all backgrounds, using coincidence signals in the Skin and OD.

\section{Deuterium-Deuterium Neutron Source Calibration} \label{dd}

Besides neutron sources deployed through the CSD, LZ also utilizes an Adelphi Technology deuterium-deuterium (DD) neutron generator~\cite{adelphiTech} to produce neutrons for TPC calibrations and for cross-checking OD tagging efficiency. For the TPC, the DD source is used for NR calibration up to the 74~keV$_{nr}$ endpoint, DAQ trigger efficiency measurements, the evaluation of the single scatter reconstruction efficiency, and the NR light and charge yields at low energies as described in reference~\cite{LZ:2022ufs}. 

\begin{figure}
    \centering \includegraphics[width=0.8\textwidth]{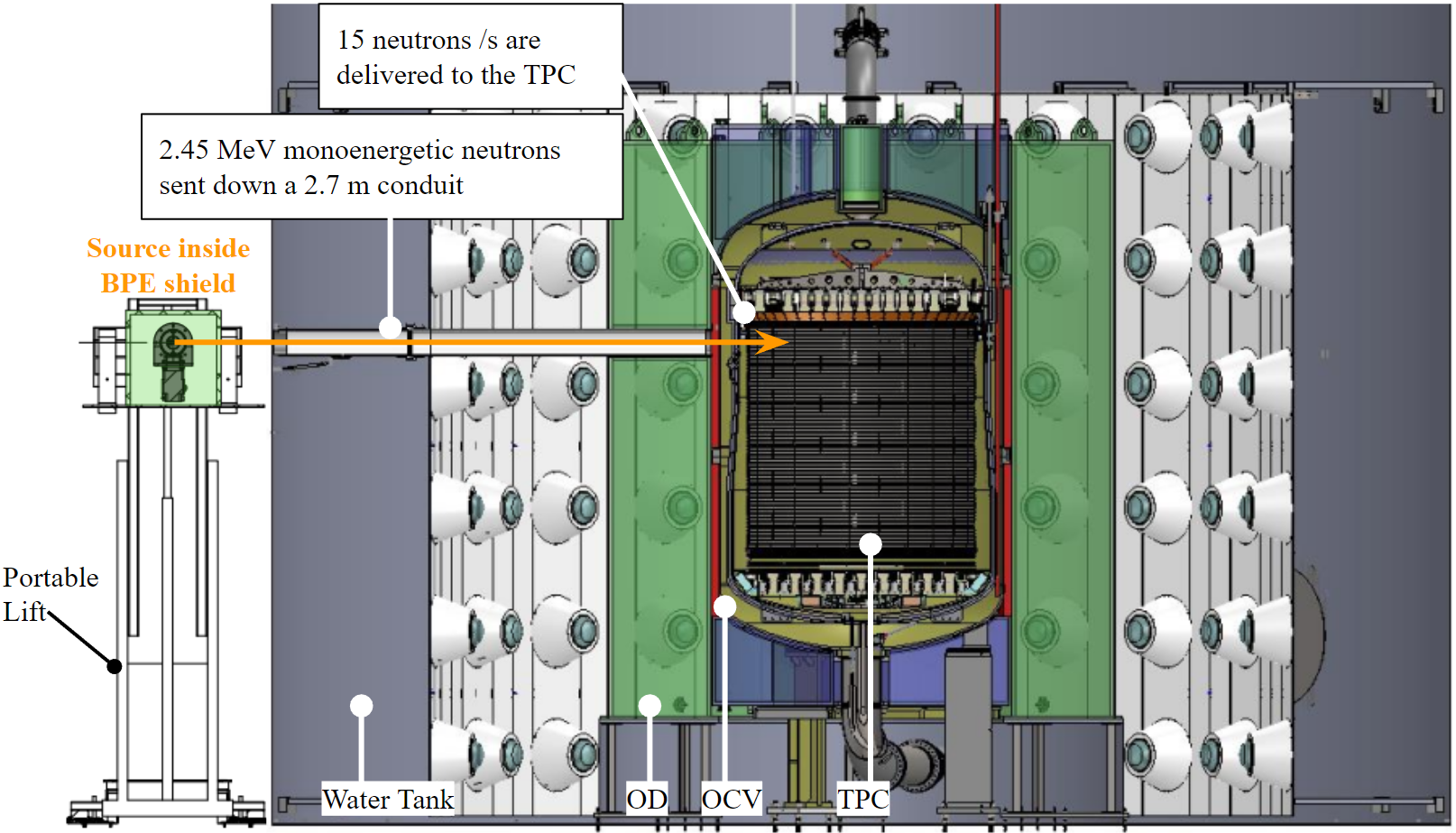}
    \label{fig:DD_cad}
  \caption{A CAD drawing of DD neutron generator deployment in the LZ experiment with respect to the horizontal conduit. In the Direct mode, monoenergetic 2.45~MeV neutrons are directly sent down the conduit. The angled conduit is 90$^\circ$ off from the horizontal conduit, which is presented in Figure~\ref{fig:dep-methods} but not shown here. During calibration, the generator surrounded by borated polyethylene (BPE) is positioned outside the water tank on a portable lift. The neutron beam is collimated by the nitrogen-purged conduits which traverses the water tank and the OD.}
  \label{fig:DD_cad_and_beam}
\end{figure}

\begin{figure}
 \centering
 \includegraphics[width=0.85\textwidth]{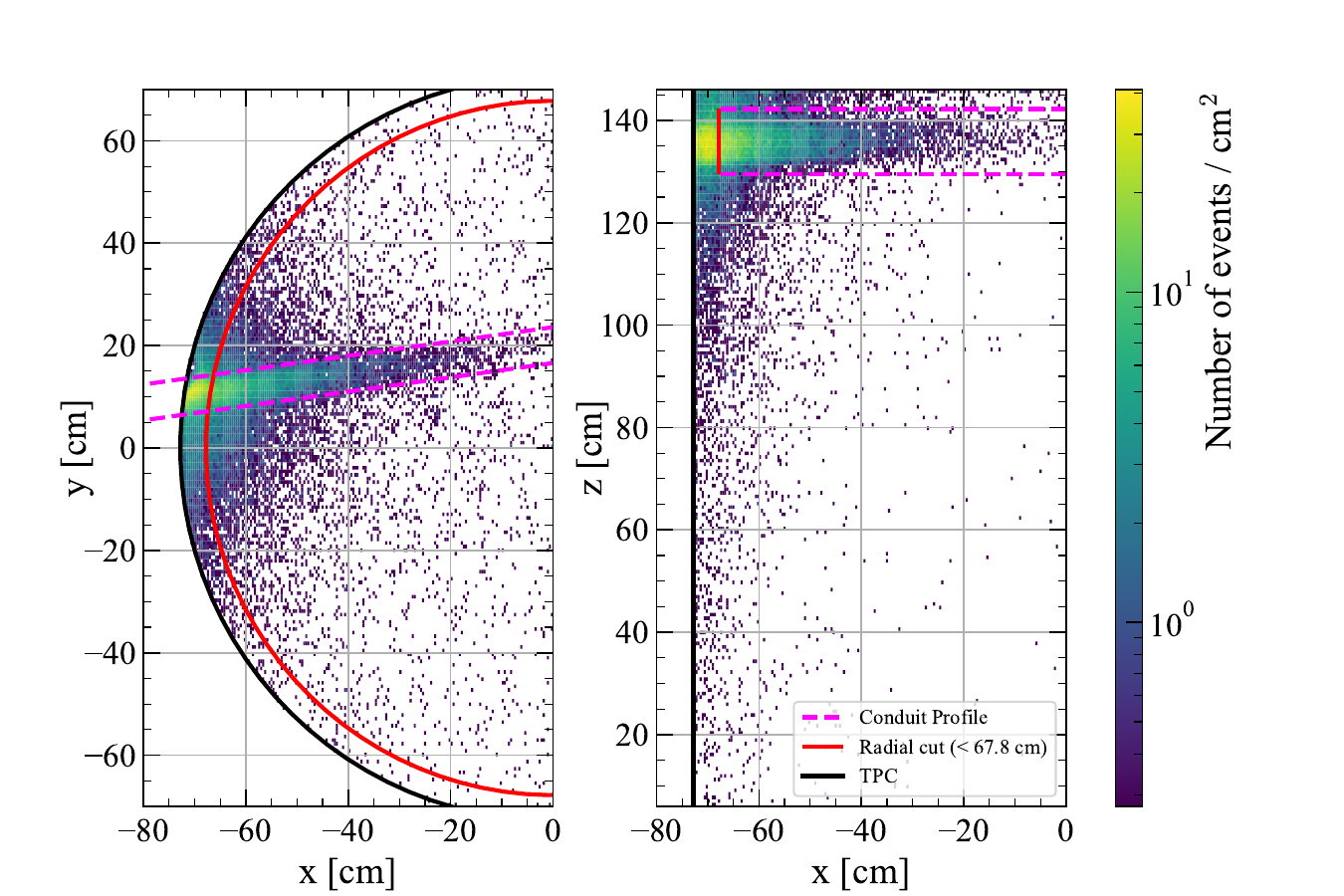}
 \caption{Reconstructed positions of events from the Direct mode DD calibration data in the Cartesian $x$, $y$, $z$ coordinates of the TPC. Events that are more than 5~cm away (dotted red line) from the TPC wall and between 129 cm and 142 cm above the cathode are used for DD neutron analysis. The neutron beam is collimated by a 4.9~cm diameter conduit and its profile is shown as the dashed purple line.}
 \label{fig:DD_neutron_beam2}
\end{figure}

The neutron production process from the DD generator is as follows: First, the deuterium is ionized and turned into plasma by a microwave (the generator has a viewing window which can be examined to check for successful plasma production). Afterwards, a high voltage is applied to a titanium-coated copper target, drawing the deuterium ions into it. Initial D$^{+}$ ions embed into the target and form titanium deuteride. Subsequent waves of D$^{+}$ ions then fuse with the embedded ions and release neutrons via the D+D$\rightarrow ^{3}$He+n reaction. The DD generator is able to produce configurable neutron energy distributions through operations in three different modes: the Direct DD, the D-Reflector, and the H-Reflector~\cite{LUX_DD, verbus_DDReflector2017, verbus_thesis,Huang_thesis,Taylor_thesis}. The DD generator is deployed outside the water tank, and neutrons are collimated and transmitted to the OCV through 2.7~m long nitrogen purged conduits (see Figures~\ref{fig:DD_cad_and_beam} and \ref{fig:DD_neutron_beam2}). There are two conduits: one is horizontal to the ground and delivers neutrons 3~cm below the liquid xenon surface inside the TPC, while the other is angled downwards at 20 degrees from the horizontal to maximize the length of neutron path inside the liquid xenon and the $z$ separation of multiple scatter events to facilitate light and charge yield measurements from NR interactions~\cite{mount2017lux}. Both conduits are made of polyvinyl chloride and are arranged in a Y-shape (see Figure~\ref{fig:dep-methods}) with a narrow branch (of 4.9~cm inner diameter) joined in the middle of a straight wider branch (of 14.6~cm inner diameter). The narrow branch is used to deliver neutrons in the Direct mode, while the wider branch is used for Reflector modes, since the rates of neutrons reaching the TPC in Reflector modes are lower. The generator itself is fixed on a portable Ekko lift~\cite{EKKO_EA15B} to facilitate its movement between the conduits and their branches. The conduits are connected to a plumbing system that enables them to be filled with deionized water (for external background shielding) when the generator is not in operation. In preparation for the DD calibration, the water in the conduits is drained by flushing it with nitrogen. The conduits experience buoyancy in the water tank during this process, so special care is taken to ensure excessive upward pressure is not applied to the acrylic tanks. We monitor OD liquid level metrics during conduit draining and filling, and performs water leak-in checks of the conduits after each DD run. The use of structural supports and careful selection of materials and geometries to dissipate forces over large areas also mitigates the risk of damage to the acrylic tanks, and are recommended for future experiments.

The Direct mode sends monoenergetic 2.45~MeV neutrons down the conduit to the TPC, directly from the neutron production surface at the center of the generator head. In the Direct mode, the neutron production is pulsed to suppress the ER background rate and enables the selection of events coincident with neutron production. Coupled with the narrow profile of the DD conduit in drift time (70~$\upmu$s wide), a stringent S2 timing cut can be applied; thereby making the DD effective in studying sub-threshold S1 events and S2-only events by reducing electron-train background noise through time tagging~\cite{collaboration2022improved}. Neutron pulsing is achieved by turning the plasma on and off within the main generator chamber. The timing of neutron production is determined by measuring plasma intensity through a viewing window on the chamber. During the SR1 calibration, the Direct mode operated at a pulse frequency of 150~Hz with a pulse width of 50~$\upmu$s. With this setup, 15 neutrons per second are delivered into the TPC. This Direct mode, characterized by high intensity, a single energy, and low ER background rate, made it a suitable source for conducting the search for the Migdal effect (a nuclear recoil interaction accompanied by atomic ionization)~\cite{ibe2018migdal}.

\begin{figure}[!ht]
 \centering
 \includegraphics[width=0.76\textwidth]{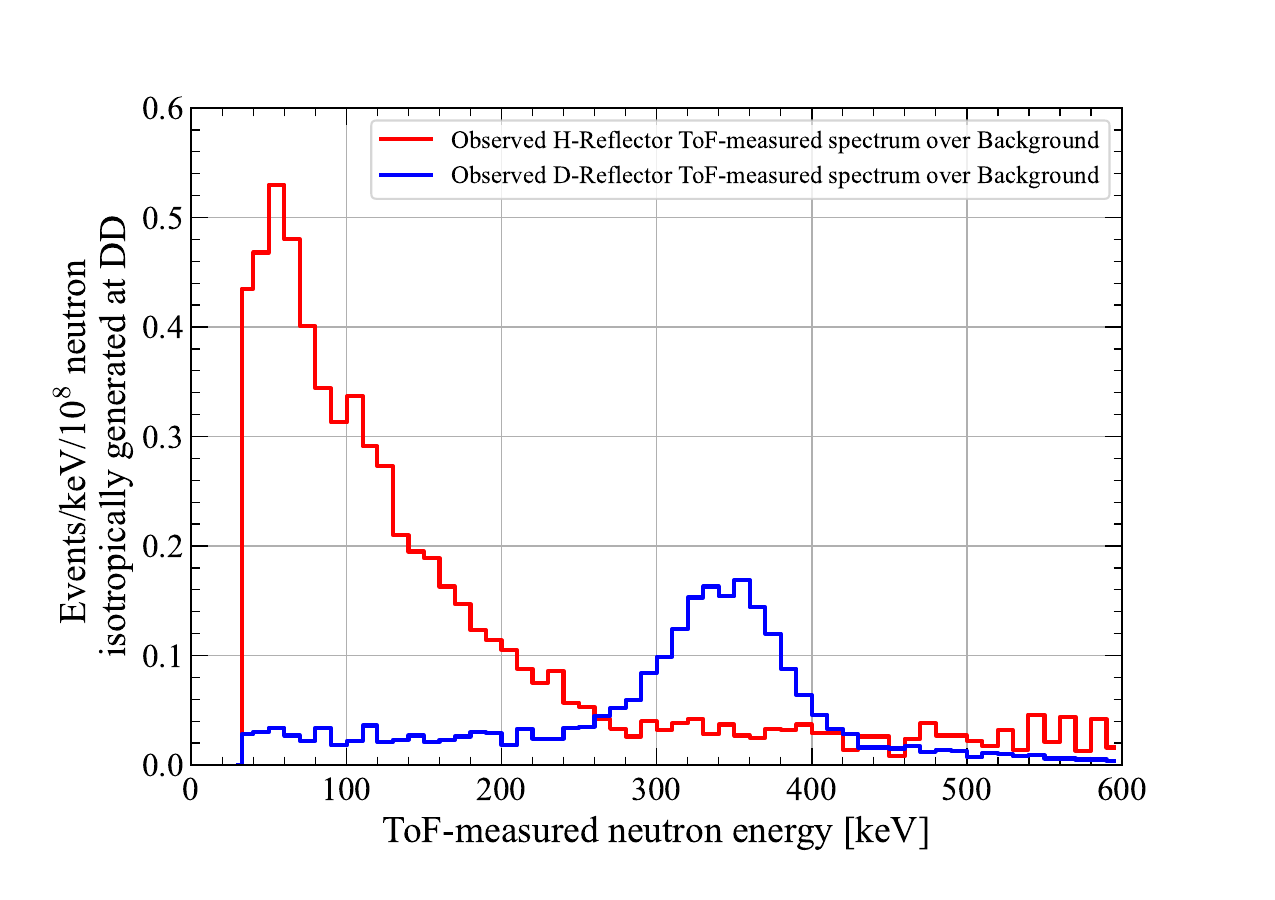}
 \caption{Measured time of flight (ToF) neutron energy of the D- (blue) and H- (red) Reflector modes in a test setup before their use in LZ.
 The energy spectrum of each Reflector mode can be tuned by the geometric configuration.} 
 \label{fig:DD_refToF}
\end{figure}

\begin{figure}[!ht]
 \centering
 \includegraphics[width=\textwidth]{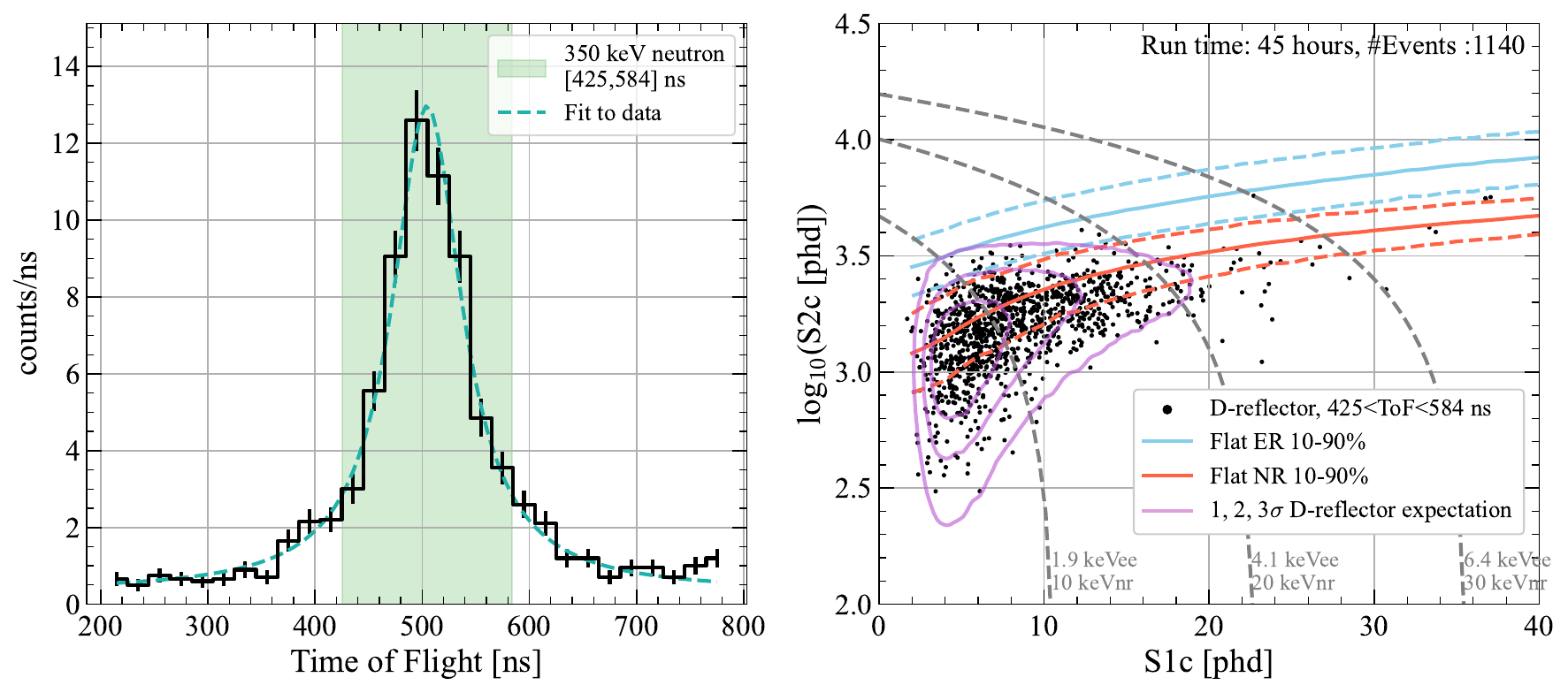}
 \caption{Left: Time of flight (ToF) distribution of neutron events from the D-Reflector calibration in the LZ experiment. The peak is at 505 $\pm$ 2~ns, which corresponds to a neutron kinetic energy of 349 $\pm$ 3~keV. This is close to the desired 350~keV neutrons with a ToF of $\sim$500~ns given the distance between the generator and the TPC. Events within the green box are selected for the analysis based on expected ToF of neutron events from the D-Reflector. Right: Detector response to D-Reflector events with ToF between 425 and 584~ns as indicated by the green box on the left plot. The solid red (blue) line is the median of a flat NR (ER) distribution and the dashed red (blue) lines are the 10-90\% quantiles modelled using $\textsc{nest}$ 2.3.7~\cite{szydagis2011nest}. The dashed grey lines are contours of constant energies. Purple contours are the expected S1 and S2 distribution of neutrons from the D-reflector. Most events are within the contours as expected.
 }
 \label{fig:DD_LZToFS1S2}
\end{figure}

The D-Reflector (H-Reflector) mode reflects neutrons from the production surface at selected scattering angles off a deuterium-based (hydrogen-based) active scintillator to send neutrons from a desired lower energy spectrum into the TPC, each with a per-neutron time-of-flight (ToF) based energy tag. The ToF is obtained by measuring the time difference between the S1 pulse obtained in the TPC and the light signal obtained from the PMT coupled to the scintillator in the D-Reflector (or H-Reflector) mode. The D-Reflector consists of a 7.6~cm diameter, 7.6~cm tall cylindrical aluminum cell filled with deuterated benzene liquid scintillator~\cite{Eljen_EJ315} coupled to a 3-inch PMT~\cite{ETEL_model_9821KEB}. The H-Reflector consists of a cuboid 10~cm $\times$ 15~cm$ ~\times~ $2.5~cm plastic scintillator~\cite{Eljen_EJ200} coupled to a 2-inch PMT~\cite{ETEL_model_9266KFLB53}. The D- and H-Reflectors are oriented with respect to the neutron generator's production surface and LZ neutron conduit to select for neutrons scattering into the LZ TPC at particular angles. 
Hence, the energy spectrum of Reflector modes can be customized by adjusting the geometric configuration.
The D-Reflector mode selects neutrons recoiling off deuterium atoms at a 135$^{\circ}$ angle, sending a peaked neutron spectrum at $349~\pm~3$~keV with FWHM of $79~\pm~2$~keV into the detector. The H-Reflector mode selects neutrons recoiling off hydrogen atoms between 78$^{\circ}$ and 85$^{\circ}$, sending a low energy (10 -- 200~keV) spectrum of ToF-tagged neutrons into the detector. Both Reflector spectra were measured using ToF tests at an LZ test facility before the DD generator was shipped to SURF, as shown in Figure~\ref{fig:DD_refToF}. 
These Reflector modes are successfully used to calibrate the low energy response of LZ down to a few keV, and the result from the D-Reflector mode measurement is shown in  Figure~\ref{fig:DD_LZToFS1S2}. This is the first time a DD Reflector mode measurement has been made in a tonne-scale liquid xenon detector. The ToF-tagged neutrons delivered by the Reflector modes with known incident energies are useful for low-energy neutron calibration down to a few keV and for efficiency measurement. Reducing incident neutron energy allows for the time separation of progressively closer separated scatters in position. The separation of the two S1 signals within a double-scatter event in the TPC enables a direct measurement of the light yield from the first energy deposition. The energy deposited in the first scatter can be reconstructed using the known (tagged) initial neutron energy and the observed scattering angle. Thus, the low-energy light and charge yield measurements are improved.

The DD Direct and the two Reflector modes each have a custom neutron shielding structure surrounding the generator, made of 5\% borated polyethylene (BPE). The purpose of the Direct mode shielding is to reduce the entry of neutrons into the cavern. The Reflector mode shielding is additionally designed to screen out all secondary-scattered neutrons from entering the LZ TPC via the conduit. In DD fusion, gamma production is suppressed, and the gamma-to-neutron ratio is $\sim$ 10$^{-7}$~\cite{DD_cecil1985measurement}. However, bremsstrahlung radiation and neutron interactions with the generator material can produce secondary x-rays. During Direct mode calibration runs, the DD generator housing is additionally surrounded by a 6~mm thick lead shield to reduce X-ray and gamma fluxes from the generator. The Reflector mode, in contrast, demands minimal material near the generator production surface. Nearby material could induce secondary scatters of higher-energy neutrons into the Reflectors or down the LZ conduits, with a similar ToF to the target low-energy neutrons. Therefore, instead of a full shield, only a 6~mm Pb plate is placed at the conduit entrance outside the water tank. A 10 inch Bonner sphere~\cite{bonner} and a plastic scintillator ~\cite{Eljen_EJ200} are mounted under the DD generator platform to continuously monitor neutron intensity during all calibration runs. Measuring the neutron intensity emitted by the generator not only provides the neutron count entering the detector for detection efficiency studies, but also serves as a safety interlock that will shut down the generator if the dose rate exceeds the radiation safety benchmark of 0.5~mrem/hour.

\section{Photoneutron Source (YBe) Calibration} \label{ybe}
Understanding the low energy detection efficiency of the LZ TPC is essential to achieving high sensitivities to low-mass (< 10~GeV) dark matter and CE$\nu$NS from solar $^8$B neutrinos~\cite{ma2023search, aprile2021search}. LZ utilizes a photoneutron source based on the ($\gamma, n$) reaction of yttrium-beryllium (YBe) \cite{collar2013applications} to calibrate nuclear recoil responses in this low energy range. The neutron energy from the YBe photonuclear reaction is 152.3$\pm$3.7~keV \cite{Knoll:2010xta}, leading to a $\sim$4.6~keV$_{nr}$ end-point recoil energy from elastic scattering on xenon nuclei.~The detector response to nuclear recoils below this energy is especially interesting as it covers the energy range of $^8$B neutrinos ($<2$~keV$_{nr}$).

\begin{figure}[ht]
\centering
\begin{minipage}{.635\textwidth}
\includegraphics[width=1\linewidth]{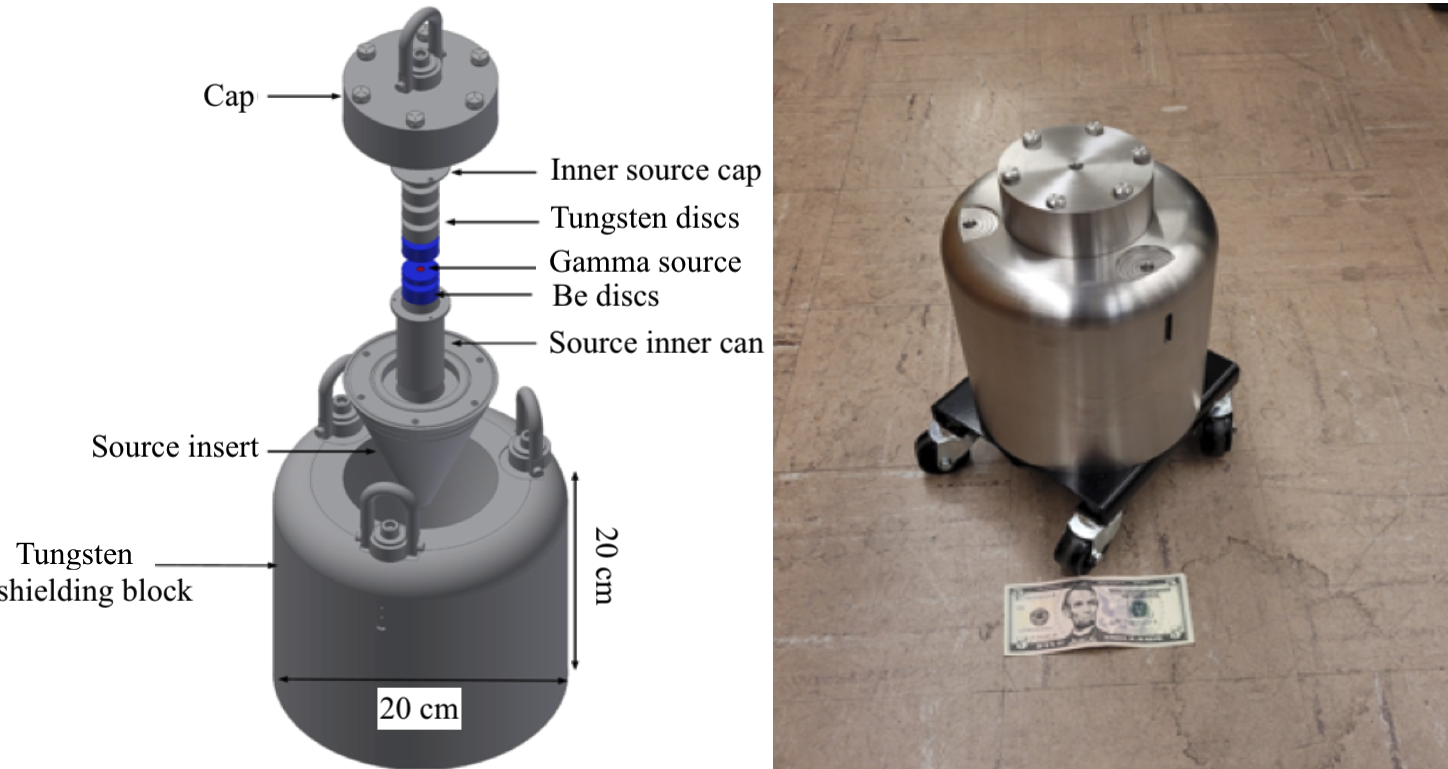}
\end{minipage}
\begin{minipage}{.33\textwidth}
\includegraphics[width=1\linewidth]{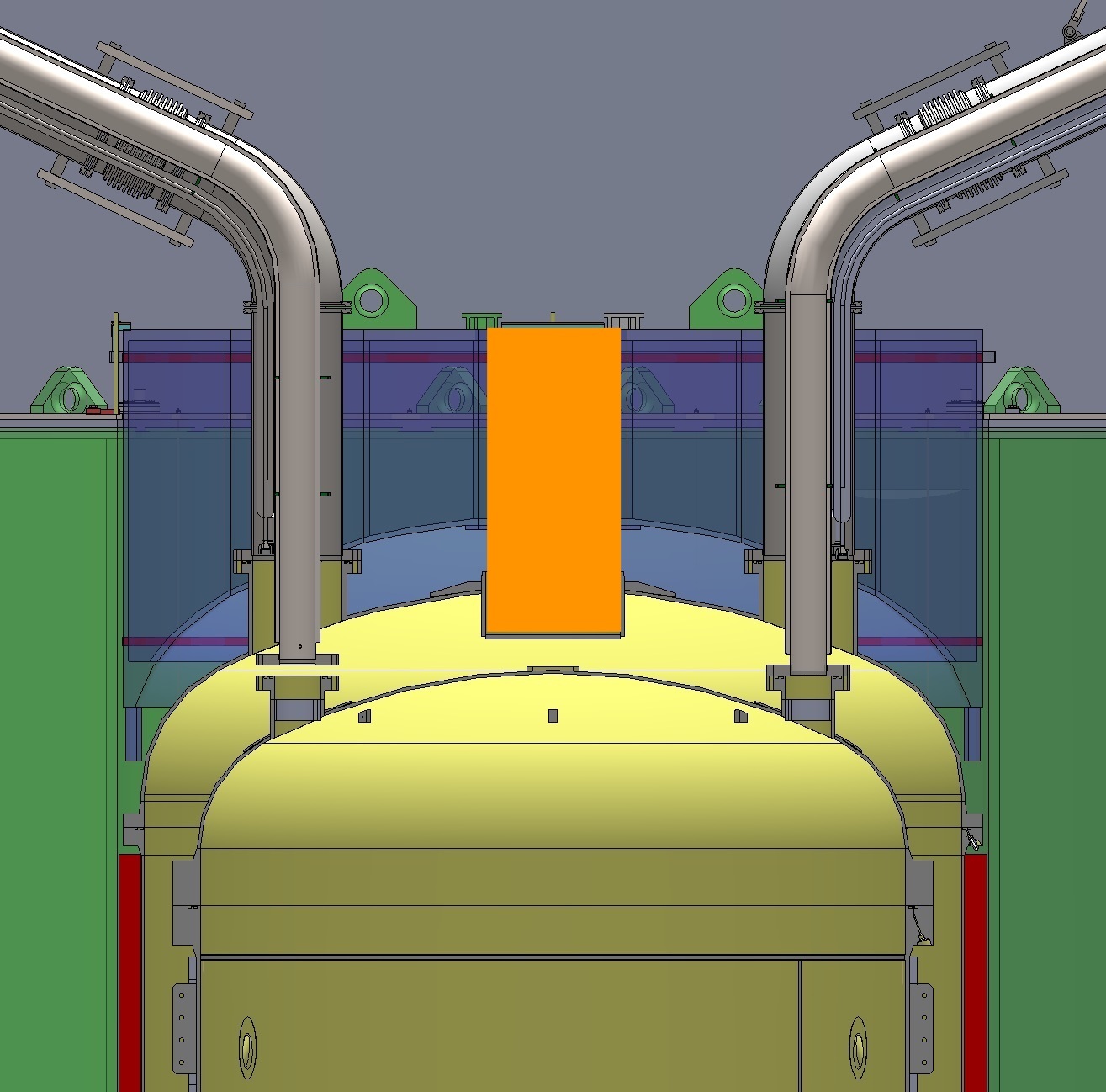}
\end{minipage}
    \caption{Left: CAD drawing for the YBe photo-neutron source assembly. The $^{88}$Y sealed source is sandwiched between $^{9}$Be disks to generate neutrons. The tungsten shielding reduces the number of $^{88}$Y $\gamma$-rays that can enter the detector. Middle: YBe source inside the tungsten shielding placed next to a five-dollar bill for scale. Right: Layout of the top part of the OD and water tank. The custom cut-out (orange) in the acrylic vessels (purple and green) through which the YBe source is deployed is shown.}
    \label{fig:YBe}
\end{figure}

\begin{figure}[!htb]
    \centering
    \includegraphics[width=0.45\linewidth]{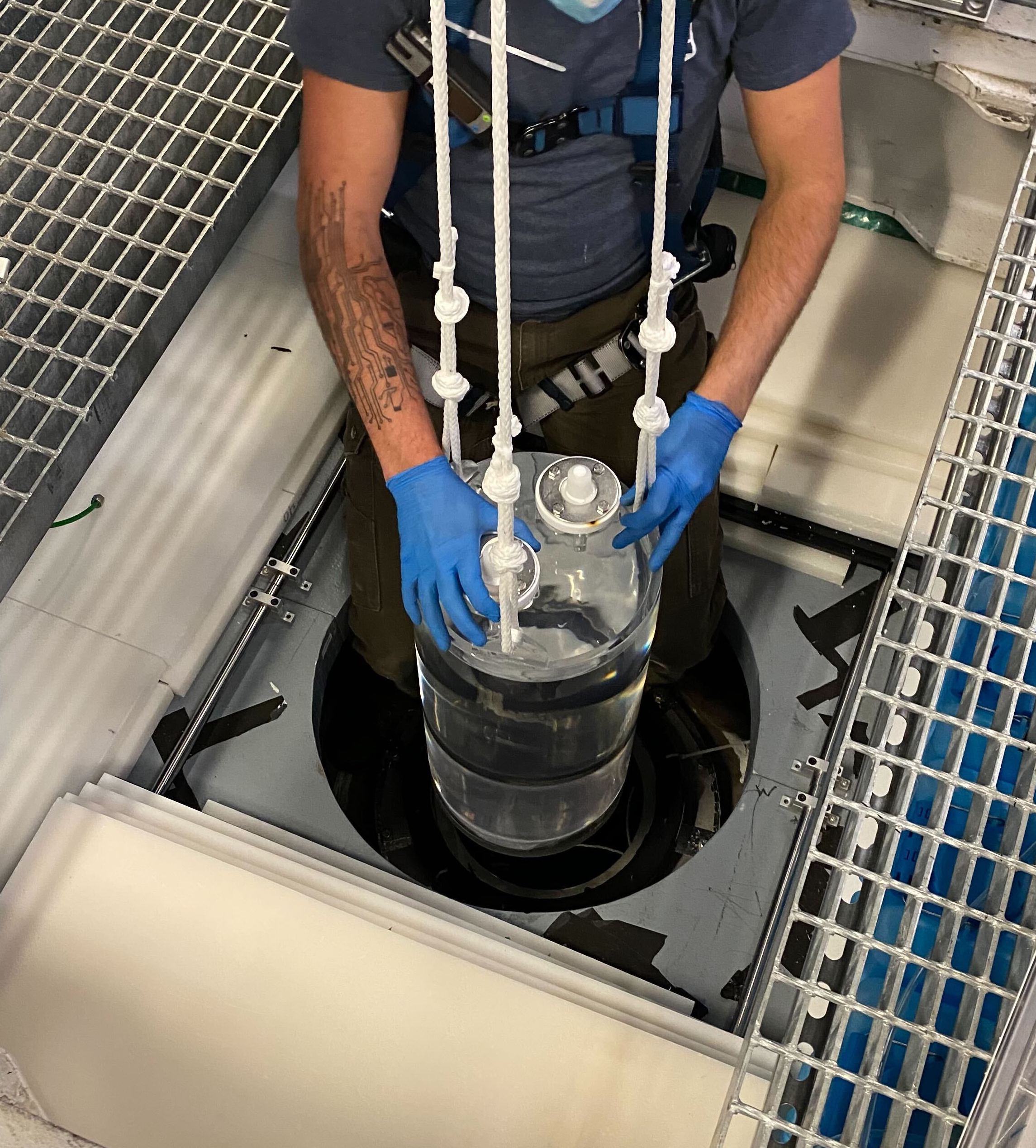}
    \caption{Deployment of the GdLS plug -- a small cylindrical acrylic tank with neutral buoyancy. It is used to fill the cut-out during the science run and is only replaced with the YBe source during calibrations.}
    \label{fig:YBeplug}
\end{figure}

The source consists of three $^{9}$Be metal disks with 24~$\upmu$m nickel plating surrounding the $^{88}$Y sealed source 
to generate neutrons, as shown in Figure~\ref{fig:YBe} (left). Only $\sim$1 neutron is produced per 10$^4$ $\gamma$-rays emitted by $^{88}$Y decays due to the small production cross-section~\cite{collar2013applications}, so gamma shielding around the YBe source is necessary to increase the neutron-to-gamma ratio entering the TPC. This shielding is provided by a tungsten cone with tungsten disks stacked on top of the $^{9}$Be disks. The tungsten cone containing the YBe source is placed inside a larger tungsten shielding block of 20~cm diameter and 20~cm height, as shown in Figure~\ref{fig:YBe} (left and middle). Two nitrile rubber O-rings between the tungsten cone and the block guarantee a water-tight seal necessary for the source deployment inside the water tank. During the calibration, the top of the water tank is opened and the YBe source assembly is deployed with a crane to the top of the outer cryostat vessel through a custom cut-out in the center of the top OD acrylic vessel (marked in orange in Figure~\ref{fig:YBe} (right)). This deployment location is chosen for mechanical stability and proximity to the TPC. After the calibration, the YBe source assembly is lifted out and replaced with an acrylic cylinder containing GdLS referred to as the ``GdLS plug''. A picture of this plug is shown in Figure~\ref{fig:YBeplug}. The GdLS plug is kept in place and swapped for the YBe source only during calibration runs. This design makes the photoneutron source as close to the TPC as possible during the calibration, while maintaining the background neutron tagging ability of the OD when the calibration campaign is finished. The water tank opening time for the YBe source deployment is minimized to mitigate against air ingress, which could cause water contamination and an OD background rate increase. The ingress can be further reduced in the future by building a high-flow nitrogen purge system to prevent air entry during the exchange of the GdLS plug and the YBe source. 

\begin{figure}[!htb]
\centering
\includegraphics[width=0.99\linewidth]{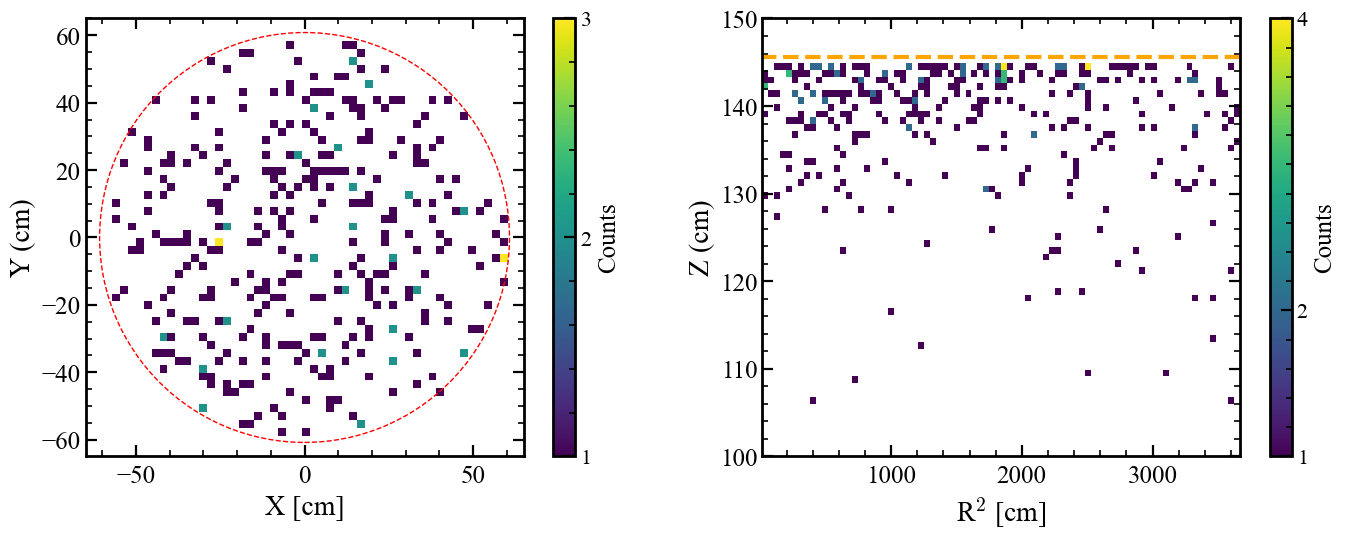}
\caption{Reconstructed positions of 200 events from the YBe calibration data in the Cartesian coordinates of the TPC. Left: $x$ vs.~$y$ distribution. The red contour in the left plot represents the radial selection cut applied to the data. Right: $z$ vs.~R$^2$ distribution. The majority of YBe events appear in the top few cm underneath the liquid-gas interface (marked by the orange line in the right plot) since neutrons from the YBe source located on top of the OCV can only penetrate a few centimeters into the liquid xenon. The gap between the liquid-gas surface and the events is the result of the drift time selection.}
\label{fig:ybe_data}
\end{figure}

In order to better understand the impact of gamma backgrounds from $^{88}$Y on the YBe data analysis, a yttrium-magnesium (YMg) gamma source is also deployed in LZ. YMg does not produce any neutrons because the excitation energy required to produce free neutrons emission in Mg is higher than the $^{88}$Y gamma energy~\cite{international2000iaea}. However, the $\gamma$-ray attenuation properties of magnesium and beryllium are the same to within $\sim$5\% \cite{NISTXCOM}, meaning the amount of $\gamma$-rays from $^{88}$Y emitted by the YMg source is similar to that from the YBe source. Shortly after the YBe calibration in LZ, beryllium metal disks were swapped out for magnesium metal disks and the tungsten block containing the YMg source was deployed to the same cut-out location in the OD. Almost no events from the YMg source were observed in the energy region of interest for neutrons emitted by the YBe source, confirming that events observed in the YBe calibration data are indeed low-energy photoneutrons produced by the source. About 200 single scatter neutrons were detected after all analysis cuts from a 112 live hour YBe run, and the measured detector response to these neutrons matches well with the expectation from simulations~\cite{collaboration2021simulations, YBepaper}. This is the first photoneutron calibration data set to have been taken in a tonne-scale detector and enabled the calibration of the $^8$B solar neutrino energy region. A dedicated publication of the YBe data analysis and its results is in preparation~\cite{YBepaper}.
The spatial distributions of the selected YBe single scatter events are shown in Figure~\ref{fig:ybe_data}. Their homogeneous radial profile is in accordance with the source deployment location, which is right above the top center of the OCV enveloping the TPC. This is in contrast to the profile of events from a CSD source, which is asymmetrical and biased toward the side tube in which the source is deployed.

\section{Detector Optical Calibrations}\label{pmtcalib}
In order to obtain accurate detector response signals induced in PMTs during the data collection described in section~\ref{disbursed} - \ref{ybe}, the PMTs themselves should be regularly calibrated to account for any time variation in their performance. This section discusses the optical PMT calibration systems in LZ, which are divided according to the volumes the PMTs monitor: the Xe PMT system (including both the TPC and Skin detectors)  and the OD PMT system. These systems focus on calibrating the TPC, Skin and OD PMT response, the PMT stability over time, and the optical properties of the OD acrylic tank and the GdLS.
\subsection{Xe LED Calibration System}\label{tpcleds}  
\begin{figure*}
    \centering
    \includegraphics[width=0.9\linewidth]{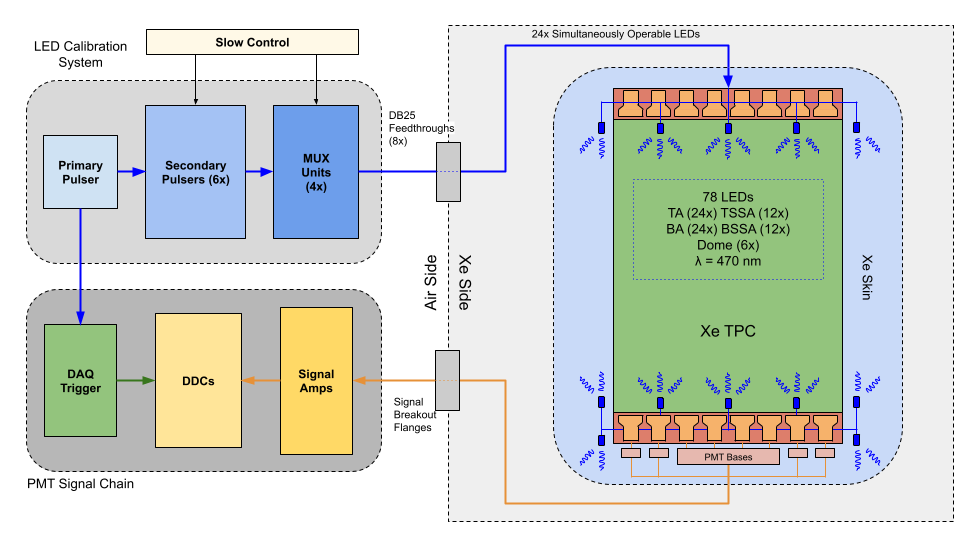}
    \caption{A diagram showing the full Xe LED Calibration System. Signal from a primary pulser can simultaneously drive up to 24 of the 78 LEDs installed throughout the TPC and Xe skin. The primary pulser also acts as an event trigger for the digital data collectors (DDCs) via the LZ data acquisition (DAQ) system. Note that the LEDs do not protrude into the TPC, and that skin and dome PMTs are excluded from this diagram.}
    \label{fig:XLCS} 
\end{figure*}

The LZ Xe PMT system consists of 494 and 131 VUV PMTs from Hamamatsu~\cite{hamamatsu_pmts} viewing the TPC and the Skin xenon volume, respectively.  The calibration of these PMTs relies on an LED system built to monitor their performance and the stability of their light response.
There are 78 LEDs through the two detectors, installed such that they properly illuminate the entirety of each detector volume. In the TPC, there are 24 LEDs mounted uniformly between PMTs in both the top and bottom arrays (totalling 48 LEDs) via through-holes in the titanium mounting structure. Each LED is enclosed in a PTFE cover to preserve the reflectivity of the PTFE lining on each array. LEDs in the Skin are further subdivided into two components, the Top Side Skin Array (TSSA) LEDs located just below the liquid/gas xenon interface outside of the TPC,  and the Bottom Side Skin Array (BSSA) LEDs looking up, as indicated in Figure~\ref{fig:XLCS}. The TSSA is fitted with twelve LEDs installed above the titanium mounting structure of the top Skin PMTs. The LEDs are pointed downwards and are covered by PTFE sheets, which helps to diffuse light into the Skin. There are twelve BSSA LEDs which are pointed upwards to calibrate the top skin PMTs. They are fixed in the PTFE housing surrounding the bottom skin PMTs. Additionally, there are six LEDs pointed downwards to calibrate the PMTs that view the LXe below the bottom PMT array, known as the "dome" region. 

The LEDs used in the Xe LED calibrations system~\cite{ELLED} emit blue light with a wavelength of 470~nm. A visible light is critical in reducing systematics by minimizing the probability of producing additional photoelectrons per incoming photon via the double photoelectron emission effect~\cite{Faham_2015}. A diagram of the full Xe LED system is shown in Figure~\ref{fig:XLCS}. The LEDs are individually driven by pulse generators~\cite{SRSDG645} as shown in the Figure. A single primary pulse generator acts to synchronize LED light emission with the LZ data acquisition system by sending a 100~ns width pulse to trigger data collection surrounding the pulse emission. A simultaneous pulse from the primary pulse generator is sent to six secondary pulse generators. Each of these six secondary pulse generators has four outputs, allowing for the simultaneous operation of 24 individual LEDs. In order to distribute these 24 pulser outputs to the entire set of 78 LEDs, the outputs of the pulsers are connected to a set of four multiplexer (MUX) units shared between the pulsers.  A single MUX unit consists of a Mainframe Switch Unit~\cite{KeysightSU} and six Multiplexer Cards~\cite{KeysightMUX}. In total, these four MUX units distribute 24 pulser inputs across 96 output channels. From these outputs, 78 channels are used for operating LEDs, and the remaining 18 are kept as spares. With independent control over each pulse generator output, up to 24 LEDs can be switched on simultaneously and each one with different pulse settings. Both the secondary pulsers and the MUX units are remotely operated through connection to the LZ slow control system, where pulse settings are chosen and distributed depending on the specific requirements of a calibration.

\begin{figure*}[ht]
    \centering
    \subfloat{{\includegraphics[width=0.5\textwidth]{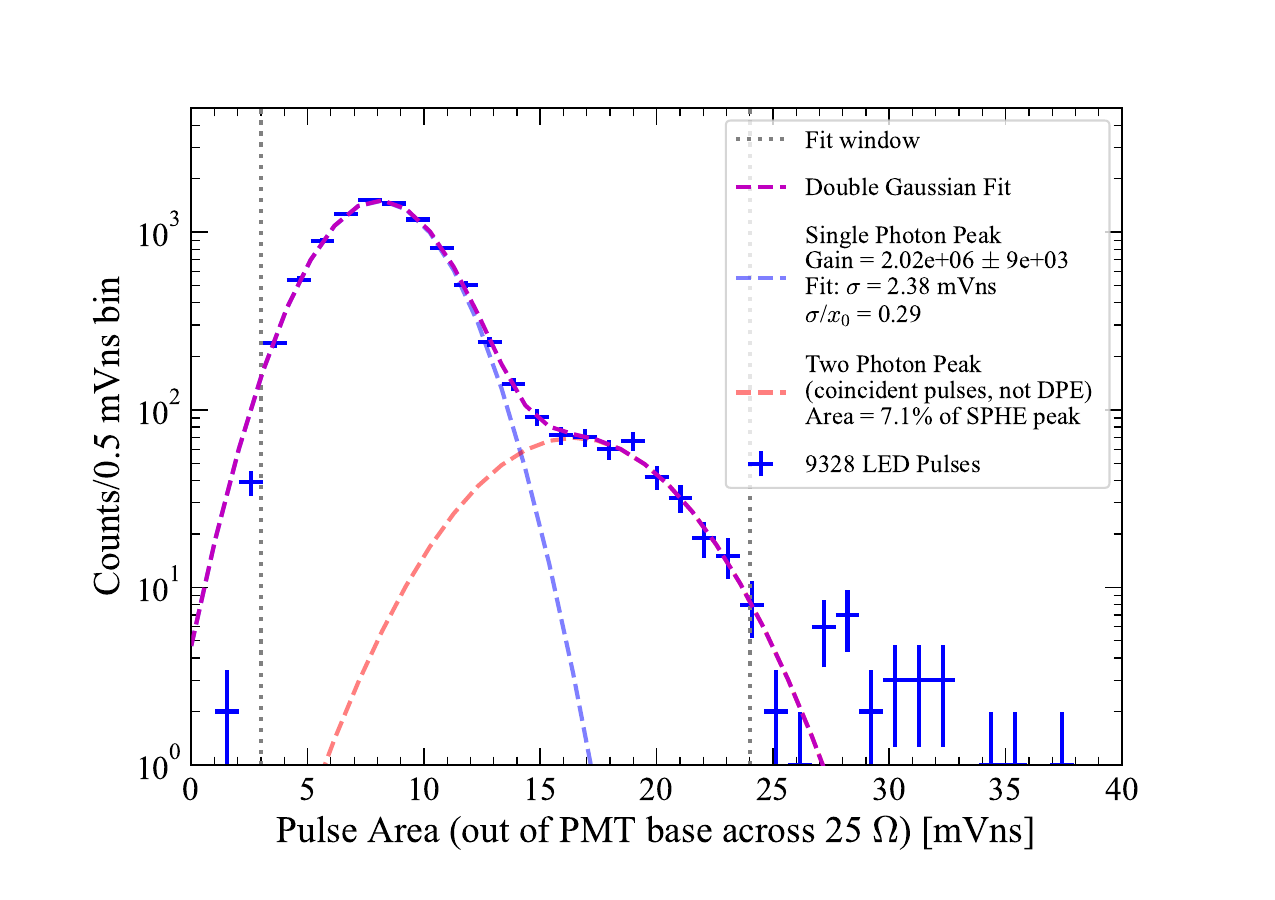} }}
    \subfloat{{\includegraphics[width=0.5\textwidth]{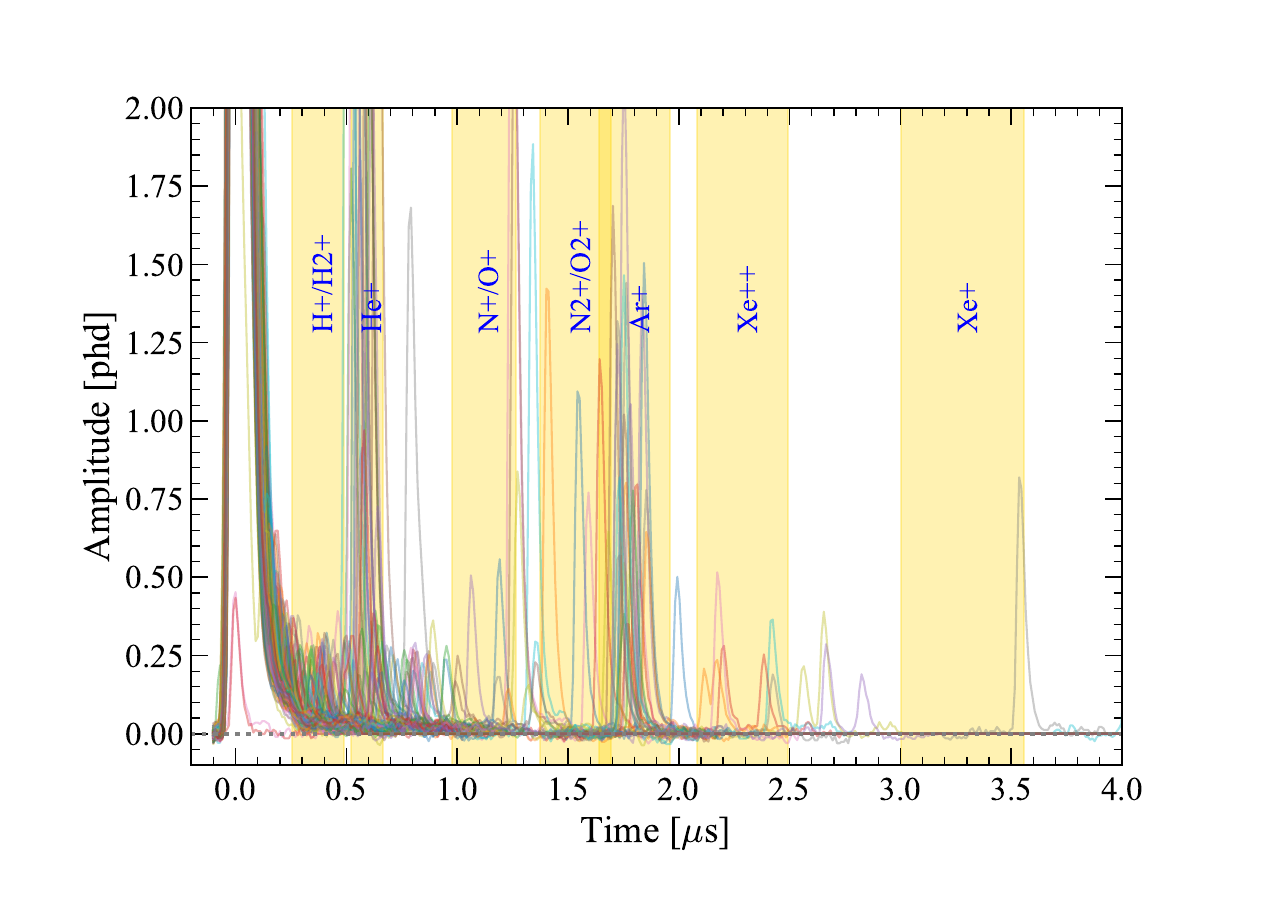} }}
    \caption{Left: A regular calibration of the Xe PMT system includes the per-PMT measurements of gain. PMT gain is determined through a double-Gaussian fit to the PhE pulse area spectrum from LED generated light. Right: Calibration of the afterpulsing ratio (APR). High intensity LED pulses are used to induce afterpulses in each PMT. Traces from many LED pulses (shown above) are co-added. APR is then measured as the ratio of the total area of each ion population to the combined area of the primary LED pulses. The timing of a single afterpulse for a particular ion is given by its mass, and shown by the yellow shaded bands. }
    \label{fig:LED_GainAPR}
\end{figure*}

The primary application of the Xe LED calibration system is to measure the response of PMTs and monitor their performance over time. LEDs are used to measure PMT gain and Afterpulsing Ratio (APR), which are a calibration of the single photoelectron (sphe) response and a measurement of contaminants in the PMT vacuum space, respectively. The PMT signal yield, or PMT gain, is normalized across all PMTs in a given system. The gain, defined as the average number of photoelectrons generated through electron multiplication after a single incoming photon, is a function of the operating voltage of the PMT. ~For the TPC PMTs, the HV bias is adjusted such that they operate at a gain of 2$\cdot10^6$ e/sphe. To calibrate the PMT gain, LEDs are driven at low amplitudes such that the observed signal is dominated by single-photons. Additionally, following a sufficiently large pulse, contaminant ions in the PMT can produce secondary ionization (afterpulses). Afterpulses are indistinguishable from standard PMT pulses and can interfere with the proper classification of events within the TPC. Proper calibration of the frequency and typical area of these afterpulses is an important prerequisite to understanding the signal output of a PMT. For afterpulsing measurements, the LEDs are driven at a higher voltage, with typical pulse areas of ~100 photons per PMT.  Example results from each measurement can be seen in Figure \ref{fig:LED_GainAPR}. For both measurements, the LEDs are pulsed at 1~kHz with a pulse width of 10~ns. In the TPC, simultaneous operation of four LEDs (two top, two bottom) is sufficient to cover all PMTs. During skin calibrations, all 24 side skin LEDs are operated simultaneously to uniformly illuminate the TSSA and BSSA PMTs. Similarly, all six dome LEDs are operated during dome PMT calibrations.

The LED system is also used to calibrate the DAQ electronics system. First, the split output of the primary pulser allows for a measurement of the timing offset between all Xe PMT signals and the event trigger.  Next, the LED system assists in validating the zero-suppression algorithm used to remove periods of data containing only baseline noise, which produces per-channel PMT waveforms of Pulse-Only Digitization (PODs). Finally, the DAQ sphe digitization efficiency is measured by determining the fraction of single photo-electrons whose signal amplitude exceeds the POD digitization threshold. This is achieved by approximating the distribution of sphe amplitudes as a Gaussian (see Figure~\ref{fig:LED_GainAPR}). Each of these calibrations is completed by generating pulsed single-photons from the LED system.

The Xe LED calibration was used extensively to monitor the long-term stability of the Xe PMTs during SR1. The average gain of the 482 operational TPC PMTs was measured to be stable within 1\% during SR1, which spanned a five month period. Periodic optical calibrations will continue into future science runs of LZ in order to track both the health of individual PMTs and the large-scale trends of the Xe PMT system.

\subsection{OD Optical Calibration System}\label{ocs}
The LZ OD PMT system consists of 120 8-inch PMTs mounted on a cylindrical array of 20 ladders inside the water tank with six PMTs on each ladder, as seen in Figure~\ref{fig:overview}. 
\begin{figure*}[ht]
    \centering
    \includegraphics[width=0.999\linewidth]{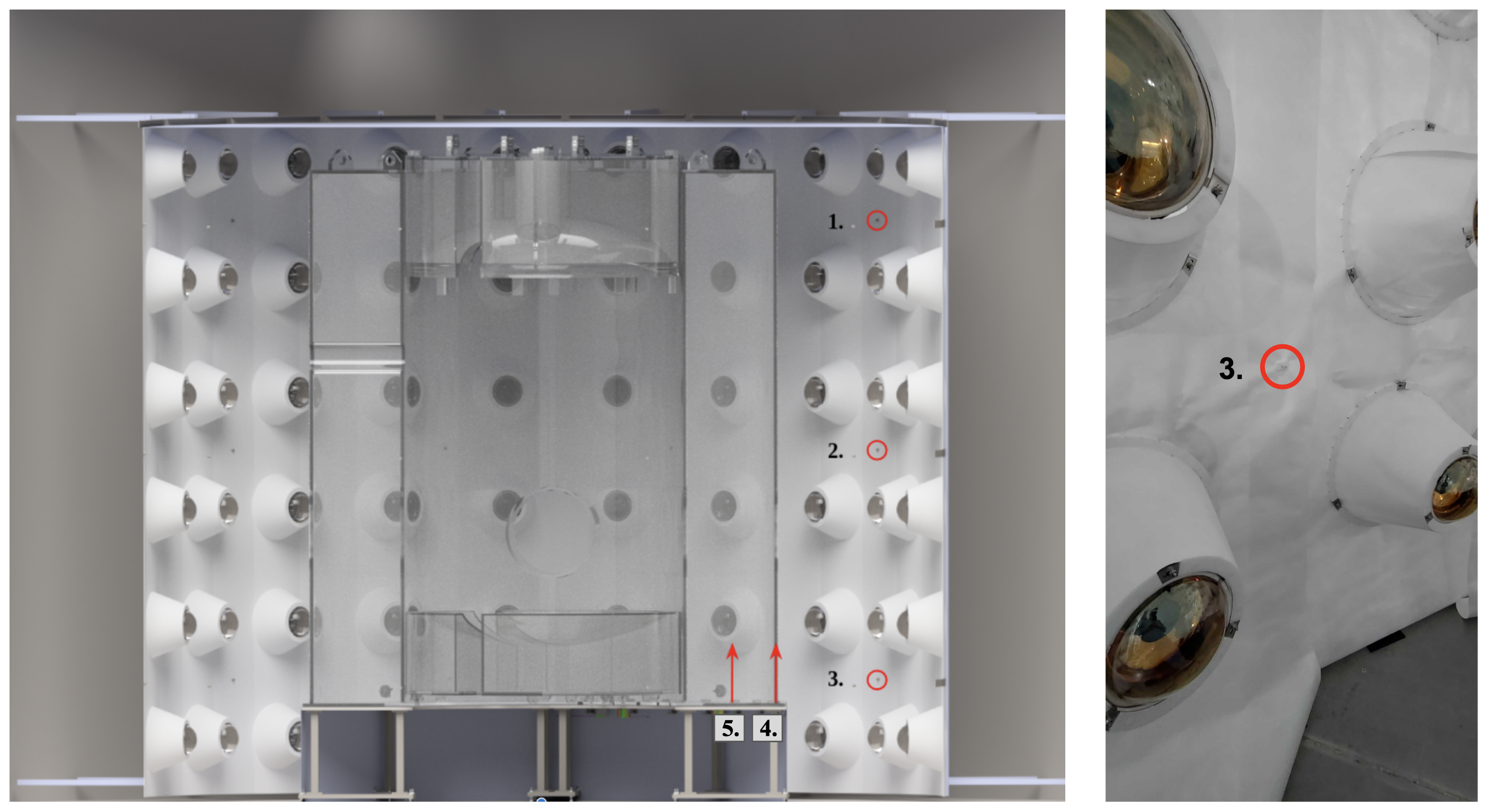}
    \caption{Left: A CAD drawing of the cross-section of the OD, showing the side acrylic tanks and the OD PMTs mounted on their ladders. The three nominal heights of the 10 azimuthal positions of the optical fiber injection points are labeled 1-3. Additionally, two injection points under the bottom acrylic tank are labeled 4 and 5. Right: A photo of the bottom OD PMT array showing one of the injection points.}
    \label{fig:OD_PMT_CAD_picture}
\end{figure*}
An Optical Calibration System (OCS), designed to monitor the optical properties of the OD down to 150~keV (which is the relevant energy range for neutron tagging with the OD), is used to calibrate the OD PMT gain/sphe response and afterpulsing. The OCS uses duplex optical fibers to inject pulses of light produced by LEDs into the OD at 35 locations. Thirty injection points are evenly distributed within the OD PMT array (10 azimuthal positions at 3 heights as shown in Figure~\ref{fig:OD_PMT_CAD_picture}). 
Five injection points are located beneath the four side acrylic tanks directing light upwards into the tanks. Four of the injection points are positioned in the center of the four acrylic tanks' base to monitor the optical properties of the liquid scintillator. The last one is located in the rim of one of the acrylic tanks and is used to monitor the optical properties of the acrylic which constitutes the vessel. For the 30 injection points situated within the PMT array, LEDs of 435~nm  are used to match the peak wavelength and quantum efficiency of the OD PMTs. Only one of the cores of the duplex fiber is used to inject light into the detector. The other core is available for potential future upgrades or in case of damage to the first core. The transmission of light through the acrylic is wavelength dependent, so 435~nm and 450~nm LEDs are used for the four injection points located beneath the scintillator tanks to monitor scintillator degradation. A 390~nm and 435~nm pair of LEDs is used to monitor the absorption of UV light and degradation of the acrylic tanks. Both cores of the duplex fibers are used for the upward facing injection points to send light in different directions. Monitoring the optical properties of the scintillator and acrylic is paramount to check if light collection during science runs is consistent.

\begin{figure*}[ht]
    \centering
    \includegraphics[width=0.9\linewidth]{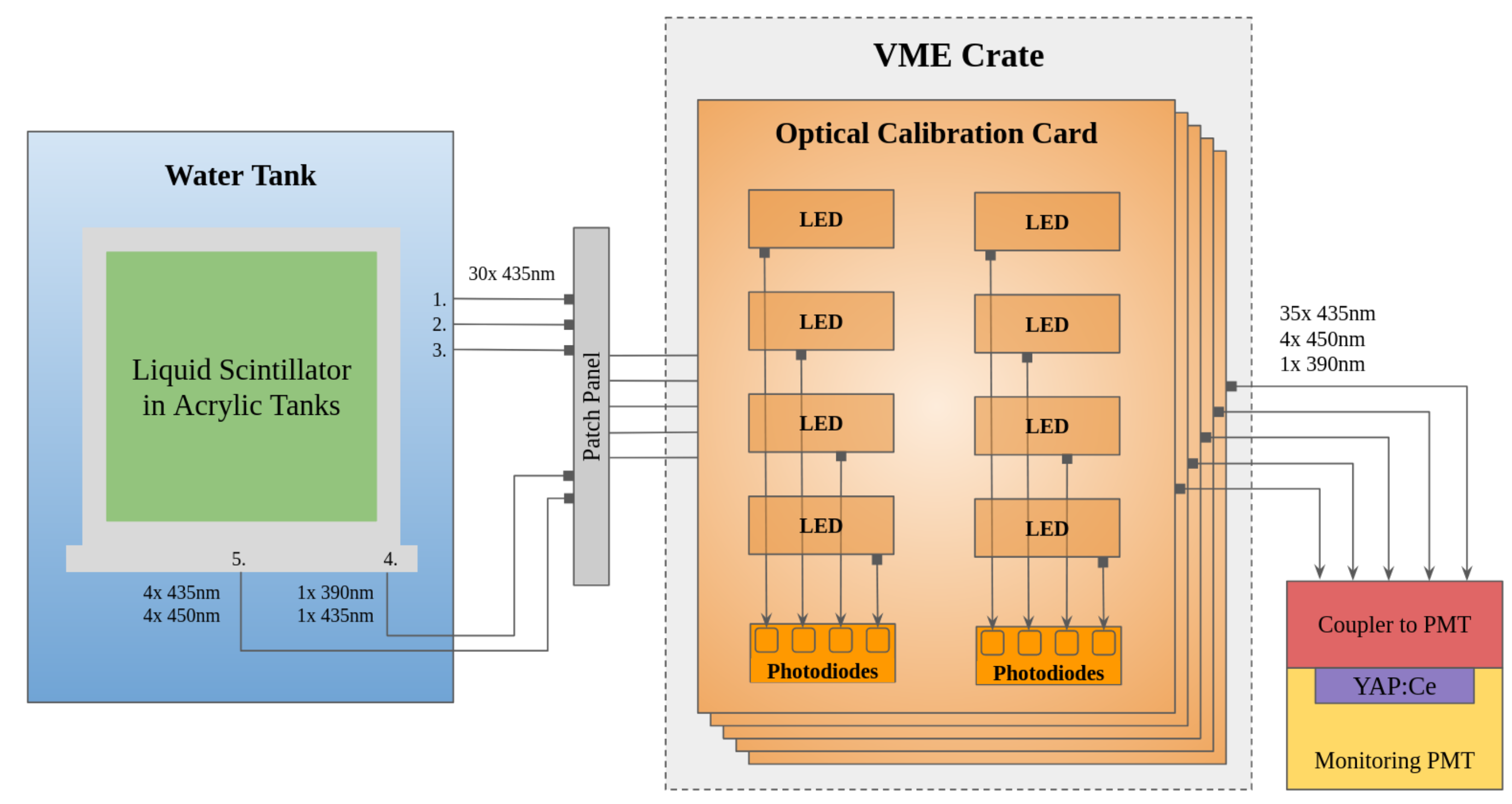}
    \caption{An overview of the OD Optical Calibration System, with eight LED pulsers on one Optical Calibration Card (OCC) and five OCCs in the VME crate. Lines with arrows show fibre routes with labels representing numbers of fibers from LEDs with corresponding wavelengths~\cite{ODOCS_NIM}. Labels 1-3 represent the three heights for the ten azimuthal positions, while Label 4-5 denote the injection points to monitor the bottom acrylic tanks, as shown in Figure~\ref{fig:OD_PMT_CAD_picture}.}
    \label{fig:OCSOverview}
\end{figure*}

The OCS electronics consist of five custom optical calibration cards.
Each card consists of an FPGA controlled motherboard housing eight custom-made LED pulser boards and two custom-made photo-diode boards. Light from the LED pulser is divided using a three-way optical coupler into a photo-diode input and two outputs on the front panel, allowing the intensity of injected light to be monitored. Light is fed from the rack housing the OCS electronics to the positions within the water tank via duplex optical fibers. The layout of the system can be seen in Figure~\ref{fig:OCSOverview}. This layout allows each individual LED to be controlled separately, with the capability to pulse multiple LEDs simultaneously to reach greater intensities of light. 

The intensity of light produced by the OCS is monitored in two ways: through the FPGA controlled photo-diode boards and via an 8-inch Hamamatsu R5912 PMT \cite{ODPMT}, which is installed in a rack mounted dark box close to the OCS electronics. This PMT is identical to those used in the OD. The stability of this PMT is also monitored, using a YAP:Ce pulser unit which produces light pulses corresponding to five thousand photo-electrons with a rate of 20~counts/s~\cite{YapCe}. The intensity of light produced by the OCS can be compared against light produced by the unit to monitor the stability of the light produced by the OCS. The OCS is controlled using the LZ slow control system, allowing a user to define a pulse configuration using a graphical user interface. 

The OCS was extensively tested during LZ commissioning and met all initial design requirements \cite{ODOCS_NIM}. It was then used to calibrate the gain/sphe response of the OD PMTs while monitoring their afterpulsing rates over time. During SR1, an average gain drift across all OD PMTs of +0.6\% was observed, demonstrating the stability of the OD system. No optical degradation (of the acrylic nor of the liquid scintillator) was observed in the OD during our continuous monitoring. 

\section{Conclusions}
In this paper, we described the technical details of the LZ calibration systems. We performed extensive quality assurance checks and performance tests on all these systems to ensure they achieve the intended science goals. 
A number of novel features and technologies implemented in the calibration systems that improved their overall performance are worth emphasizing.

We developed a dispersed source injection system with precise dose control. It enables the injection of a wide range of gaseous source activity into the detector through the circulation system. A CH$_4$ purifier installed in the dispersed source injection system allows the removal of gaseous species containing tritium that can stay on cold detector surfaces for a long time. The use of this purifier ensures safe tritium calibrations and avoids issues such as unexpected amount of residual tritium observed in other experiments~\cite{PhysRevLett.129.161805,PhysRevLett.127.261802}. 

Facilitated by a laser feedback system, the CSD system for deploying external rod sources to different $z$-positions in three calibration tubes, is able to achieve mm-precision.  Using the CSD gamma sources, we were able to calibrate the energy scale and inter-detector timing among TPC, Skin, and OD for particle interactions taking place at various locations. We designed and manufactured novel AmLi neutron sources with low neutron emission rates that are not commercially available. These AmLi sources are used to calibrate the detector response to nuclear recoils and measure neutron tagging efficiency of the OD. 

A DD neutron generator with high neutron-to-gamma ratio and configurable neutron energy via Direct and Reflector modes was implemented in LZ. It was the first use of neutron Reflector modes in a large scale detector calibration. Moreover, a novel, custom-made photon-neutron (YBe) source was deployed for the first time in a noble liquid dark matter experiment to calibrate the detector response to neutrons with energy depositions below 4.6~keV$_{nr}$ in LXe. The calibration data obtained from this source is critical for future measurements of the coherent scattering of $^{8}$B solar neutrinos with xenon nuclei in the LZ detector. 

We also implemented a state-of-the-art optical calibration system to monitor the TPC, Skin and OD PMT gains and their stability over time, as well as the optical properties of the OD acrylic tank and the GdLS. This is to ensure an accurate understanding of the detector response signals induced in the PMTs in these three different volumes during data collection; enhancing LZ background rejection capability and its signal discovery power. 

Overall, the design and implementation of the hardware documented in this paper as well as the calibration data presented therein, are crucial for the published and future LZ science results. They also provide important guidance for future calibrations in rare-event searching experiments employing similar detector technologies.

\begin{acknowledgments}
The research supporting this work took place in part at SURF in Lead, South Dakota. Some of the isotopes used in this research was supplied by the U.S. Department of Energy Isotope Program, managed by the Office of Isotope R$\&$D and Production.~Funding for this work is supported by the U.S. Department of Energy, Office of Science, Office of High Energy Physics under Contract Numbers DE-AC02-05CH11231, DE-SC0020216, DE- SC0012704, DE-SC0010010, DE-AC02-07CH11359, DE-SC0012161, DE-SC0015910, DE-SC0014223, DE-SC0010813, DE-SC0009999, DE-NA0003180, DE-SC0011702, DE-SC0010072, DE-SC0015708, DE-SC0006605, DE-SC0008475, DE-SC0019193, 
DE-FG02-10ER46709, UW PRJ82AJ, DE-SC0013542, DE-AC02-76SF00515, DE-SC0018982, DE-SC0019066, DE-SC0015535, DE-SC0019319, 
DE-AC52-07NA27344, DOE-SC0012447, DE-SC0024225, \& DE-SC0024114. This research was also supported by U.S. National Science Foundation (NSF); the UKRI’s Science \& Technology Facilities Council under award numbers ST/M003744/1, ST/M003655/1, ST/M003639/1, ST/M003604/1, ST/M003779/1,ST/M003469/1, ST/M003981/1, ST/N000250/1, ST/N000269/1, ST/N000242/1, ST/N000331/1, ST/N000447/1, ST/N000277/1, ST/N000285/1,
ST/S000801/1, ST/S000828/1, ST/N000739/1, ST/S000879/1, ST/S000933/1, ST/N000844/1, ST/S000747/1, ST/S000666/1, ST/R003181/1; Portuguese 
Foundation for Science and Technology (FCT) under award numbers PTDC/FIS-PAR/2831/2020; the Institute for Basic Science, Korea (budget number IBS-R016-D1). We acknowledge additional support from the STFC Boulby Underground Laboratory in the U.K., the GridPP~\cite{Faulkner_2006,ukgrid} and IRIS Collaborations, in particular at Imperial College London and additional support by the University College London (UCL) Cosmoparticle Initiative, and by the ARC Centre of Excellence for Dark Matter Particle Physics, and the University of Zurich. We acknowledge additional support from the Center for the Fundamental Physics of the Universe, Brown University. K.T. Lesko acknowledges the support of Brasenose College and Oxford University. The LZ Collaboration acknowledges key contributions of Dr. Sidney Cahn, Yale University, in the production of calibration sources. This research used resources of the National Energy Research Scientific Computing Center, a DOE Office of Science User Facility supported by the Office of Science of the U.S. Department of Energy under Contract No. DE-AC02- 05CH11231. We gratefully acknowledge support from GitLab through its GitLab for Education Program. The University of Edinburgh is a charitable body, registered in Scotland, with the registration number SC005336. The assistance of SURF and its personnel in providing physical access and general logistical and technical support is acknowledged. We acknowledge the South Dakota Governor’s office, the South Dakota Community Foundation, the South Dakota State University Foundation, and the University of South Dakota Foundation for use of xenon. We also acknowledge the University of Alabama for providing xenon. For the purpose of open access, the authors have applied a Creative Commons Attribution (CC BY) licence to any Author Accepted Manuscript version arising from this submission.
\end{acknowledgments}

\bibliographystyle{JHEP}
\bibliography{ref}
\end{document}

%% file: LZ_Latest_Author_List-copy_to_Overleaf_Elsevier_April24_2024.tex
\author[1,2]{J.~Aalbers}
\author[1,2]{D.S.~Akerib}
\author[3]{A.K.~Al Musalhi}
\author[3]{F.~Alder}
\author[4,5]{C.S.~Amarasinghe}
\author[1,2]{A.~Ames}
\author[1,2]{T.J.~Anderson}
\author[6]{N.~Angelides}
\author[6]{H.M.~Ara\'{u}jo}
\author[7]{J.E.~Armstrong}
\author[1,2]{M.~Arthurs}
\author[6]{A.~Baker}
\author[8]{S.~Balashov}
\author[9]{J.~Bang}
\author[5,40]{E.E.~Barillier}
\author[4]{J.W.~Bargemann}
\author[10]{K.~Beattie}
\author[11]{T.~Benson}
\author[7]{A.~Bhatti}
\author[12,10]{A.~Biekert}
\author[1,2]{T.P.~Biesiadzinski}
\author[5,40]{H.J.~Birch}
\author[13]{E.~Bishop}
\author[14]{G.M.~Blockinger}
\author[15]{B.~Boxer}
\author[8]{C.A.J.~Brew}
\author[16]{P.~Br\'{a}s}
\author[17]{S.~Burdin}
\author[1,2]{M.~Buuck}
\author[18]{M.C.~Carmona-Benitez}
\author[17]{M.~Carter}
\author[19]{A.~Chawla}
\author[10]{H.~Chen}
\author[11]{J.J.~Cherwinka}
\author[18]{Y.T.~Chin}
\author[20]{N.I.~Chott}
\author[21]{M.V.~Converse}
\author[3]{A.~Cottle}
\author[22]{G.~Cox}
\author[22]{D.~Curran}
\author[23,24]{C.E.~Dahl}
\author[3]{A.~David}
\author[22]{J.~Delgaudio}
\author[25]{S.~Dey}
\author[18]{L.~de~Viveiros}
\author[6]{L.~Di Felice}
\author[9]{C.~Ding}
\author[26]{J.E.Y.~Dobson}
\author[21]{E.~Druszkiewicz}
\author[27]{S.R.~Eriksen}
\author[1,2]{A.~Fan}
\author[25]{N.M.~Fearon}
\author[25]{N.~Fieldhouse}
\author[10]{S.~Fiorucci}
\author[27]{H.~Flaecher}
\author[17]{E.D.~Fraser}
\author[28]{T.M.A.~Fruth}
\author[9]{R.J.~Gaitskell}
\author[22]{A.~Geffre}
\author[20]{J.~Genovesi}
\author[3]{C.~Ghag}
\author[12,10]{R.~Gibbons}
\author[29]{S.~Gokhale}
\author[25]{J.~Green}
\author[8]{M.G.D.van~der~Grinten}
\author[20]{J.J.~Haiston}
\author[7]{C.R.~Hall}
\author[1,2]{S.~Han}
\author[9]{E.~Hartigan-O'Connor}
\author[10]{S.J.~Haselschwardt}
\author[5,40]{M.A.~Hernandez}
\author[30]{S.A.~Hertel}
\author[5]{G.~Heuermann}
\author[31]{G.J.~Homenides}
\author[22]{M.~Horn}
\author[5,32]{D.Q.~Huang}
\author[25]{D.~Hunt}
\author[6]{E.~Jacquet}
\author[3]{R.S.~James\footnote{Also at The University of Melbourne, School of Physics, 701 Melbourne, VIC 3010, Australia}}
\author[15]{J.~Johnson}
\author[19]{A.C.~Kaboth}
\author[32]{A.C.~Kamaha}\emailAdd{akamaha@physics.ucla.edu} 
\author[14]{M.~Kannichankandy}
\author[21]{D.~Khaitan}
\author[8]{A.~Khazov}
\author[3]{I.~Khurana}
\author[4]{J.~Kim}
\author[35]{Y.D.~Kim}
\author[15]{J.~Kingston}
\author[9]{R.~Kirk}
\author[10,18]{D.~Kodroff }
\author[5]{L.~Korley}
% ...
\emailAdd{qingxia@lbl.gov}
\author[29]{X.~Xiang}
\author[36]{J.~Xu}
\author[29]{M.~Yeh}
\author[32]{E.A.~Zweig}

\affiliation[1]{SLAC National Accelerator Laboratory, Menlo Park, CA 94025-7015, USA}

\affiliation[2]{Kavli Institute for Particle Astrophysics and Cosmology, Stanford University, Stanford, CA  94305-4085 USA}

\affiliation[3]{University College London (UCL), Department of Physics and Astronomy, London WC1E 6BT, UK}

\affiliation[4]{University of California, Santa Barbara, Department of Physics, Santa Barbara, CA 93106-9530, USA}

\affiliation[5]{University of Michigan, Randall Laboratory of Physics, Ann Arbor, MI 48109-1040, USA}

\affiliation[6]{Imperial College London, Physics Department, Blackett Laboratory, London SW7 2AZ, UK}

\affiliation[7]{University of Maryland, Department of Physics, College Park, MD 20742-4111, USA}

\affiliation[8]{STFC Rutherford Appleton Laboratory (RAL), Didcot, OX11 0QX, UK}

\affiliation[9]{Brown University, Department of Physics, Providence, RI 02912-9037, USA}

\affiliation[10]{Lawrence Berkeley National Laboratory (LBNL), Berkeley, CA 94720-8099, USA}

\affiliation[11]{University of Wisconsin-Madison, Department of Physics, Madison, WI 53706-1390, USA}

\affiliation[12]{University of California, Berkeley, Department of Physics, Berkeley, CA 94720-7300, USA}

\affiliation[13]{University of Edinburgh, SUPA, School of Physics and Astronomy, Edinburgh EH9 3FD, UK}

\affiliation[14]{University at Albany (SUNY), Department of Physics, Albany, NY 12222-0100, USA}

\affiliation[15]{University of California, Davis, Department of Physics, Davis, CA 95616-5270, USA}

\affiliation[16]{{Laborat\'orio de Instrumenta\c c\~ao e F\'isica Experimental de Part\'iculas (LIP)}, University of Coimbra, P-3004 516 Coimbra, Portugal}

\affiliation[17]{University of Liverpool, Department of Physics, Liverpool L69 7ZE, UK}

\affiliation[18]{Pennsylvania State University, Department of Physics, University Park, PA 16802-6300, USA}

\affiliation[19]{Royal Holloway, University of London, Department of Physics, Egham, TW20 0EX, UK}

\affiliation[20]{South Dakota School of Mines and Technology, Rapid City, SD 57701-3901, USA}

\affiliation[21]{University of Rochester, Department of Physics and Astronomy, Rochester, NY 14627-0171, USA}

\affiliation[22]{South Dakota Science and Technology Authority (SDSTA), Sanford Underground Research Facility, Lead, SD 57754-1700, USA}

\affiliation[23]{Northwestern University, Department of Physics \& Astronomy, Evanston, IL 60208-3112, USA}

\affiliation[24]{Fermi National Accelerator Laboratory (FNAL), Batavia, IL 60510-5011, USA}

\affiliation[25]{University of Oxford, Department of Physics, Oxford OX1 3RH, UK}

\affiliation[26]{King's College London, }

\affiliation[27]{University of Bristol, H.H. Wills Physics Laboratory, Bristol, BS8 1TL, UK}

\affiliation[28]{The University of Sydney, School of Physics, Physics Road, Camperdown, Sydney, NSW 2006, Australia}

\affiliation[29]{Brookhaven National Laboratory (BNL), Upton, NY 11973-5000, USA}

\affiliation[30]{University of Massachusetts, Department of Physics, Amherst, MA 01003-9337, USA}

\affiliation[31]{University of Alabama, Department of Physics \& Astronomy, Tuscaloosa, AL 34587-0324, USA}

\affiliation[32]{University of California, Los Angeles, Department of Physics \& Astronomy, Los Angeles, CA 90095-1547, USA}

% ...
\affiliation[35]{IBS Center for Underground Physics (CUP), Yuseong-gu, Daejeon, Korea}

\affiliation[36]{Lawrence Livermore National Laboratory (LLNL), Livermore, CA 94550-9698, USA}

% ...
\affiliation[40]{University of Zurich, Department of Physics, University of Zurich, 8057 Zurich, Switzerland}